\documentclass[12pt]{iopart}
\usepackage{pdflscape}
\usepackage{epsfig}
\usepackage{iopams}
\usepackage{color}

\renewcommand{\theequation}{\thesection.\arabic{equation}}
\newcommand{\be}{\begin{equation}}
\newcommand{\ee}{\end{equation}}

\newcommand{\ba}{\begin{array}}
\newcommand{\ea}{\end{array}}
\newcommand{\bea}{\begin{eqnarray}}
\newcommand{\eea}{\end{eqnarray}}

\newcommand{\sdir}{\ensuremath{\rlap{\raisebox{.15ex}{$\mskip
6.5mu\scriptstyle+ $}}\supset}}

\newtheorem{proposition}{Proposition}

\begin{document}

\title[Fermionic supersymmetric extension of the Gauss-Codazzi equations]{Fermionic supersymmetric extension of the Gauss-Weingarten and Gauss-Codazzi equations}
\author{S Bertrand$^1$, A M Grundland$^{2,3}$ and A J Hariton$^2$}

\address{$^1$ Department of Mathematics and Statistics, Universit\'e de Montr\'eal,\\ Montr\'eal CP 6128 (QC) H3C 3J7, Canada}
\address{$^2$ Centre de Recherches Math\'ematiques, Universit\'e de Montr\'eal,\\ Montr\'eal CP 6128 (QC) H3C 3J7, Canada}
\address{$^3$ Department of Mathematics and Computer Science, Universit\'e du Qu\'ebec, Trois-Rivi\`eres, CP 500 (QC) G9A 5H7, Canada}
\ead{bertrans@crm.umontreal.ca, grundlan@crm.umontreal.ca, hariton@crm.umontreal.ca}

\begin{abstract} A fermionic supersymmetric extension  is established for the Gauss-Weingarten and Gauss-Codazzi equations describing conformally parametrized surfaces immersed in a Grassmann superspace. An analysis of this extension is performed using a superspace-superfield formalism together with a supersymmetric version of a moving frame on a surface. In contrast with the bosonic supersymmetric extension, the equations of the fermionic supersymmetric Gauss-Codazzi model resemble the form of the classical equations. Next, a superalgebra of Lie point symmetries of these equations is determined and a classification of the one-dimensional subalgebras of this superalgebra into conjugacy classes is presented. The symmetry reduction method is used to obtain group-invariants, orbits and reduced systems for three chosen one-dimensional subalgebras. The explicit solutions of these reduced systems correspond to different surfaces immersed in a Grassmann superspace. Within this framework for the supersymmetric version of the Gauss-Codazzi equations a geometrical interpretation of the results is dicussed.

\paragraph{}Keywords: supersymmetric models, Lie superalgebras, symmetry reduction, conformally parametrized surfaces. 
\end{abstract}

\maketitle

\section{Introduction}
In the last three decades, a number of supersymmetric (SUSY) extensions of classical and quantum mechanical models, describing several physical phenomena, have been developed and group-invariant solutions of these SUSY systems have been found (e.g. \cite{BJ01}-\cite{JP00}). Recently, this method was further generalized to encompass hydrodynamic-type systems (see e.g. \cite{Das}-\cite{GH11}). Their SUSY extensions were established and their group-invariant solutions were constructed. Supersymmetric versions of the Chaplygin gas in (1+1) and (2+1) dimensions were formulated by R. Jackiw et al., derived from parametrizations of the action for a superstring and a Nambu-Goto membrane, respectively (see \cite{Jackiw} and references therein). It was suggested that a quark-gluon plasma may be described by non-Abelian fluid mechanics \cite{TJZW}. In addition, SUSY extensions have been formulated for a number of soliton equations \cite{Chaichian}, including among others the Korteweg de Vries equation \cite{MathieuLabelle}-\cite{Mathieu}, the Kadomtsev-Petviashvili equation \cite{Manin}, the Sawada-Kotera equation \cite{Tian} and the sine-Gordon and sinh-Gordon equations \cite{Aratyn}-\cite{Siddiq06}.

Despite the progress made in the investigation of nonlinear SUSY systems, this area of mathematics does not yet have as solid a theoretical foundation as the classical theory of differential equations. 
This is related primarily to the fact that, due to the nature of Grassmann variables, the principle of superposition of solutions obtained from the method of characteristics cannot be applied to nonlinear SUSY systems.
In most cases, analytic methods for solving quasilinear SUSY systems of equations lead to the construction of classes of solutions that are more restricted than the general solution. One can attempt to construct more restricted classes of solutions which depend on some arbitrary functions and parameters by requiring that the solutions be invariant under certain group properties of the original system. The main advantages of the group properties appear when group analysis makes it possible to construct regular algorithms for finding certain classes of solutions without referring to any additional considerations but proceeding directly from the given system of partial differential equations (PDEs). A systematic computational method for constructing the group of symmetries of a given system of PDEs has been developed by many authors (see e.g. \cite{Clarkson},\cite{Olver}) and a broad review of recent developments in the SUSY case can be found in several books (e.g. J. F. Cornwell \cite{Cornwell}, D. S. Freed \cite{Freed}, V. Kac \cite{Kac}, V. S. Varadarajan \cite{Varadarajan} and B. De Witt \cite{DeWitt}). The methodological approach adopted in this paper is based on the symmetry reduction method (SRM) of PDEs invariant under a SUSY Lie group of point transformations. By a symmetry group of a SUSY system of PDEs, we mean a local SUSY Lie group $G$ transforming both the independent and dependent variables of the considered SUSY system of equations in such a way that a Lie supergroup transforms given solutions of the system to new solutions. The Lie superalgebra of such a group is represented by vector fields and their prolongation structures. The standard algorithms for determining the symmetry algebra of a system of equations and classifying its subalgebras have been extended in order to deal with our SUSY models (see e.g. \cite{Kac},\cite{RH90},\cite{Winternitz}).

Recent studies of the geometric properties of surfaces associated with holomorphic and nonholomorphic solutions of the SUSY bosonic Grassmann sigma models have been performed \cite{DHZ}-\cite{Witten}. A gauge invariant formulation of these SUSY models in terms of orthogonal projectors allows one to obtain explicit solutions and consequently to study the geometry of their associated surfaces. To pursue this research further, it is convenient to formulate a fermionic SUSY extension of the Gauss-Weingarten (GW) and Gauss-Codazzi (GC) equations for conformally parametrized surfaces immersed in a Grassmann superspace. A similar analysis of surfaces was performed \cite{BGH} using a formalism of a superspace and bosonic vector superfields together with a supersymmetric version of a moving frame on a surface. The bosonic SUSY extension of the GC equations was given by six equations and this formulation allowed us to discuss in detail a geometric characterization of surfaces, including their fundamental forms and their Gaussian and mean curvatures, and consequently to establish a SUSY version of the Bonnet theorem for surfaces immersed in the superspace $\mathbb{R}^{(1,1\vert2)}$. The results obtained for the SUSY systems were so promising that it seemed worthwhile to try to apply the above geometric approach, already used for a SUSY extension of the GW and GC equations involving bosonic superfields \cite{BGH}, to a similar extension of the SUSY GC equations built in terms of fermionic superfields. This is, in short, the aim of this paper.

The paper is organized as follows. In section 2, we present certain basic notions and properties of Grassmann algebras and Grassmann variables and introduce the notation that will be used throughout this paper. In section 3, we review relevant points from Lie's theory of symmetry groups for the GW and GC equations and apply an algorithm to isolate integrable systems and their associated surfaces. In section 4, we construct and investigate the SUSY extension of the GW and GC equations. 
We show that, in contrast with the previously considered bosonic SUSY extension of the GC equations \cite{BGH}, the femionic SUSY extension closely resembles the form of the classical GW and GC equations for moving frames on surfaces appearing in differential geometry. In section 5, we examine some geometric aspects of conformally parametrized SUSY surfaces. In section 6, a Lie superalgebra of symmetries of the SUSY GC equations is determined. In section 7, a systematic classification of the one-dimensional subalgebras of the Lie superalgebra into conjugacy classes is performed. Section 8 contains three examples of invariant solutions of the SUSY GC equations obtained by the SRM. Finally, in section 9, we present some conclusions and possibilities for future research.

\section{Preliminaries on certain aspects of Grassmann algebras}\setcounter{equation}{0}
The basic notions and definitions used in this section, such as the properties of Grassmann algebras and Grassmann variables, together with the introduced notation, will be used subsequently in what follows. A more complete description can be found in severals references (see e.g. \cite{Cornwell}-\cite{DeWitt},\cite{Berezin}-\cite{Rogers80}). For a brief review, we refer the reader to \cite{BGH}, section 3. In this paper we make use of a complex Grassmann algebra $\Lambda$ involving a number of Grassmann generators $(\xi_1,\xi_2,\xi_3,...)$. The actual number of Grassmann generators of $\Lambda$ does not matter as long as there are enough of them to make all formulas encountered meaningful. Any fermionic (odd) variables $\theta^+$ and $\theta^-$ satisfy the relation
\be
(\theta^+)^2=(\theta^-)^2=\theta^+\theta^-+\theta^-\theta^+=0.
\ee
The Grassmann algebra $\Lambda$ can be decomposed into the form
\be
\Lambda=\Lambda_{even}+\Lambda_{odd},
\ee
where $\Lambda_{even}$ includes all terms involving a product of an even number of generators $\xi_k$, i.e. $1,\xi_1\xi_2,\xi_1\xi_3,...$, while $\Lambda_{odd}$ includes all terms involving a product of an odd number of generators $\xi_k$, i.e. $\xi_1,\xi_2,\xi_3,...,\xi_1\xi_2\xi_3,...$ Also, the Grassmann algebra $\Lambda$ can be decomposed as
\be
\Lambda=\Lambda_{body}+\Lambda_{soul},
\ee
where $\Lambda_{body}$ includes all terms that do not include any of the generators $\xi_k$, i.e $\Lambda_{body}$ is isomorphic to $\mathbb{C}$, while $\Lambda_{soul}$ includes all terms that include at least one generator $\xi_k$. If $h$ and $g$ are Grassmann quantities, then the partial derivatives involving odd variables, $\theta^+$ and $\theta^-$, satisfy the following Leibniz rule
\be
\partial_{\theta^\pm}(hg)=(\partial_{\theta^\pm}h)g+(-1)^{\deg(h)}h(\partial_{\theta^\pm}g),
\ee
where
\be
\deg(h)=\left\lbrace\ba{l}
0\mbox{ if }h\mbox{ is even,}\\
1\mbox{ if }h\mbox{ is odd.}\\
\ea\right.
\ee
If $f$ is a function involving the variables $\theta^+$ and $\theta^-$ then we use the notation
\be
f_{\theta^+\theta^-}=\partial_{\theta^-}(\partial_{\theta^+}f).
\ee
The partial derivatives with respect to the odd Grassmann variables $\theta^+$ and $\theta^-$ satisfy
\be
\partial_{\theta^j}\theta^k=\delta_j^k,
\ee
where $\delta_j^k$ is the Kronecker delta function, and the values 1 and 2 of the indices $j$ and $k$ stand for $+$  and $-$, respectively. If $\theta$ is a fermionic-valued Grassmann variable, then the derivative $\partial_\theta$ will have the following effect on Grassmann-valued functions. If $f$ is a bosonic function, then $\partial_\theta f$ is fermionic. Similarly, if $f$ is a fermionic function, then $\partial_\theta f$ is bosonic. The interchangeability of mixed derivatives (with proper respect to the ordering of odd variables) is assumed throughout.

\section{Symmetries of the Gauss-Weingarten equations compared with symmetries of the Gauss-Codazzi equations}\setcounter{equation}{0}
Consider the GC equations for conformally parametrized surfaces immersed in 3-dimensional Euclidean space $\mathbb{R}^3$ for the unknown functions $H$, $Q$ and $e^u$
\be
\ba{ll}
\partial\bar{\partial} u+\frac{1}{2}H^2e^u-2\vert Q\vert e^{-u}=0, & (\mbox{the Gauss equation}) \\
\partial\bar{Q}=\frac{1}{2}e^{u}\bar{\partial}H,\quad \bar{\partial}Q=\frac{1}{2}e^u\partial H, & (\mbox{the Codazzi equations})
\ea\label{claGC}
\ee
whose zero curvature condition (ZCC), with potential matrices $V_1$ and $V_2$ taking values in a Lie algebra, takes the form
\be
\bar{\partial}V_1-\partial V_2+[V_1,V_2]=0,\label{claZCC}
\ee
where we have used
\be
\partial\equiv \partial_z=\frac{1}{2}(\partial_x-i\partial_y),\quad \bar{\partial}\equiv\partial_{\bar{z}}=\frac{1}{2}(\partial_x+i\partial_y),
\ee
which are the partial derivatives with respect to the complex variables $z=x+iy$ and $\bar{z}=x-iy$, respectively. The bracket $[\cdot,\cdot]$ denotes the commutator. Here, summation over repeated indices $(i=1,2,3)$ is understood and is used in what follows unless otherwise specified (by the use of a parenthesis). The conformally parametrized surface with the vector-valued function $F=(F_1,F_2,F_3)^T:\mathcal{R}\rightarrow\mathbb{R}^3$ (where $\mathcal{R}$ is a Riemann surface) satisfies the following normalization for the tangent vectors $\partial F$ and $\bar{\partial}F$, and the unit normal $N$
\be\ba{ll}
\langle\partial F,\partial F\rangle=\langle\bar{\partial} F,\bar{\partial} F\rangle=0, & \langle \partial F,\bar{\partial}F\rangle=\frac{1}{2}e^u,\\
\langle \partial F,N\rangle=\langle \bar{\partial} F,N\rangle=0, & \langle N,N\rangle=1.
\ea\ee
The coefficients $Q$, $\bar{Q}$ and $H$ of the GC equations (\ref{claGC}) involving the second derivatives of $F$ are defined as
\be
\hspace{-1cm}Q=\langle\partial^2F,N\rangle\in\mathbb{C},\quad \bar{Q}=\langle\bar{\partial}^2F,N\rangle\in\mathbb{C},\quad H=2e^{-u}\langle\partial\bar{\partial}F,N\rangle\in\mathbb{R}.
\ee
The bracket $\langle\cdot,\cdot\rangle$ denotes the scalar product in Euclidean space $\mathbb{R}^3$
\be
\langle a,b\rangle=a^ib_i.
\ee
The matrix representation (\ref{claZCC}) is the compatibility condition of a matrix differential linear system for the wavefunction $\Phi$ taking values in the corresponding Lie group $GL(3,\mathbb{C})$
\be\hspace{-2.5cm}\ba{cc}
\partial\Phi=V_1\Phi, & \bar{\partial}\Phi=V_2\Phi,\\
V_1=\left(\ba{ccc}
\partial u & 0 & \frac{1}{\lambda}Q \\
0 & 0 & \frac{\lambda}{2}He^u \\
-\lambda H & -\frac{2}{\lambda}e^{-u}Q & 0
\ea\right), & V_2=\left(\ba{ccc}
0 & 0 & \frac{1}{2\lambda}He^u \\
0 & \bar{\partial}u & \lambda\bar{Q} \\
-2\lambda e^{-u}\bar{Q} & -\frac{1}{\lambda}H & 0
\ea\right),
\ea\label{LP}\ee
where $\lambda\in\mathbb{C}$ is a unitary constant, $\vert\lambda\vert=1$. Note that we have inserted a free parameter $\lambda$ into this linear system. The linear equations (\ref{LP}) describing the kinematics of the moving frame $\Phi=(\partial F,\bar{\partial}F,N)^T$ associated with a surface when $\lambda=1$ are known as the GW equations and are identified as the non-parametric linear problem in differential geometry
\be\ba{l}
\partial\left(\ba{c}
\partial F\\
\bar{\partial}F\\
N
\ea\right)=\left(\ba{ccc}
\partial u & 0 & Q \\
0 & 0 & \frac{1}{2}He^u \\
-H & -2e^{-u}Q & 0
\ea\right)\left(\ba{c}
\partial F\\
\bar{\partial}F\\
N
\ea\right),\\
 \\
\bar{\partial}\left(\ba{c}
\partial F\\
\bar{\partial}F\\
N
\ea\right)=\left(\ba{ccc}
0 & 0 & \frac{1}{2}He^u \\
0 & \bar{\partial}u & \bar{Q} \\
-2e^{-u}\bar{Q} & -H & 0
\ea\right)\left(\ba{c}
\partial F\\
\bar{\partial}F\\
N
\ea\right).
\ea\label{claGW}\ee
The GC equations (\ref{claGC}) are the necessary and sufficient conditions for the existence of conformally parametrized surfaces in $\mathbb{R}^3$ and, according to the Bonnet theorem \cite{Bonnet}, the immersion function $F$ is unique up to Euclidean motions in $\mathbb{R}^3$. The symmetry algebra $\mathcal{L}$ of the GW equations (\ref{claGW}) is spanned by the following vector fields
\be
\ba{l}
X(\eta)=\eta(z)\partial_z-\eta'(z)(U\partial_U+2Q\partial_Q),\\
Y(\zeta)=\zeta(\bar{z})\partial_{\bar{z}}-\zeta'(\bar{z})(U\partial_U+2\bar{Q}\partial_{\bar{Q}}),\\
e_0=-H\partial_H+Q\partial_Q+\bar{Q}\partial_{\bar{Q}}+2U\partial_U+F_i\partial_{F_i},\\
T_i=\partial_{F_i},\qquad i=1,2,3\\
D_i=F_i\partial_{F_{(i)}}+N_i\partial_{N_{(i)}},\\
R_{ij}=(F_i\partial_{F_j}-F_j\partial_{F_i})+(N_i\partial_{N_j}-N_j\partial_{N_i}),\qquad i< j=2,3\\
S_{ij}=(F_i\partial_{F_j}+F_j\partial_{F_i})+(N_i\partial_{N_j}+N_j\partial_{N_i}),
\ea\label{clavf}
\ee
where we have used the notation $U=e^u$, $\eta'(z)=d\eta/dz$ and $\zeta'(\bar{z})=d\zeta/d\bar{z}$. Here $\eta$ and $\zeta$ are arbitrary functions of $z$ and $\bar{z}$, respectively.

 The symmetry algebra $\mathcal{L}$ spanned by the infinitesimal generators given in (\ref{clavf}) include two infinite-dimensional subalgebras generated by $X(\eta)$ and $Y(\zeta)$. The vector fields $e_0,D_1$ and $D_2$ correspond to three types of dilations, the $T_i$ generate translations in the $F_i$ directions respectively, $R_{ij}$ is the rotation in the $F_i$ and $N_i$ variables and $S_{ij}$ is a local boost transformation. The symmetry algebra $\mathcal{L}'$ of the linear system (\ref{LP}) is spanned by the generators $X(\eta)$ and $Y(\zeta)$ of equation (\ref{clavf}) together with a modified version of the generator $e_0$, given by
\be
e_0=-H\partial_H+Q\partial_Q+\bar{Q}\partial_{\bar{Q}}+2U\partial_U.
\ee
It should be noted that the parameter $\lambda$ does not appear in the latter symmetry algebra $\mathcal{L}'$. The commutation relations of the algebra $\mathcal{L}$ are
\be
\hspace{-2.5cm}\ba{l}
[X(\eta_1),X(\eta_2)]=(\eta_1 '\eta_2-\eta_1\eta_2 ')\partial_z+(\eta_1 ''\eta_2-\eta_1\eta_2 '')(U\partial_U+2Q\partial_Q),\\{}
[Y(\zeta_1),Y(\zeta_2)]=(\zeta_1 '\zeta_2-\zeta_1\zeta_2 ')\partial_{\bar{z}}+(\zeta_1 ''\zeta_2-\zeta_1\zeta_2 '')(U\partial_U+2\bar{Q}\partial_{\bar{Q}}),\\{}
\hspace{-0.2cm}\ba{lll}[X(\eta),Y(\zeta)]=0,& [X(\eta),e_0]=0,& [Y(\zeta),e_0]=0,\\{}
[X(\eta),T_i]=0,& [Y(\zeta),T_i]=0,& [e_0,T_i]=-T_i,\\{}
[X(\eta),D_i]=0,& [Y(\zeta),D_i]=0,& [e_0,D_i]=0,\\{}
[X(\eta),R_{ij}]=0,& [Y(\zeta),R_{ij}]=0,& [e_0,R_{ij}]=0,\\{}
[X(\eta),S_{ij}]=0,& [Y(\zeta),S_{ij}]=0,& [e_0,S_{ij}]=0,\\{}
[T_i,D_j]=\delta_{ij}T_i,& [T_i,R_{jk}]=\delta_{ij}T_k-\delta_{ik}T_j,& [T_i,S_{jk}]=\delta_{ij}T_k+\delta_{ik}T_j,\ea\\{}
[D_i,R_{jk}]=\delta_{ij}S_{ik}-\delta_{ik}S_{ij},\qquad [D_i,S_{jk}]=\delta_{ij}R_{ik}-\delta_{ik}R_{ji},\\{} [R_{ij},S_{kl}]=\delta_{jk}S_{il}+\delta_{jl}S_{ik}-\delta_{ik}S_{jl}-\delta_{il}S_{jk}.
\ea\ee
Since the vector fields $X(\eta)$ and $Y(\zeta)$ form an Abelian algebra, the vector fields (\ref{clavf}) determine that the algebra $\mathcal{L}$ can be decomposed as a direct sum of two infinite-dimensional Lie algebras together with a 13-dimensional subalgebra generated by $e_0,T_i,D_i,R_{ij}$ and $S_{ij}$, i.e.
\be
\mathcal{L}=\lbrace X(\eta)\rbrace\oplus\lbrace Y(\zeta)\rbrace\oplus\lbrace e_0,T_i,D_i,R_{ij},S_{ij}\rbrace.
\ee
This algebra represents a direct sum of two copies of the Virasoro algebra together with the 13-dimensional algebra generated by $e_0,T_i,D_i,R_{ij}$ and $S_{ij}$. Assuming that the functions $\eta$ and $\zeta$ are analytic in some open subset $\mathcal{D}\subseteq\mathbb{C}$, we can develop $\eta$ and $\zeta$ as power series with respect to $z$ and $\bar{z}$, and the results provide a basis for $\mathcal{L}$. The largest finite-dimensional subalgebra $L$ of the algebra $\mathcal{L}$ is spanned by the following $19$ generators
\be
\hspace{-1cm}\ba{l}
\hat{e}_0=-H\partial_H+Q\partial_Q+\bar{Q}\partial_{\bar{Q}}+2U\partial_U+F_i\partial_{F_i},\\
e_1=\partial_z,\quad e_3=z\partial_z-2Q\partial_Q-U\partial_U,\quad e_5=z^2\partial_z-4zQ\partial_Q-2zU\partial_U,\\
e_2=\partial_{\bar{z}},\quad e_4=\bar{z}\partial_{\bar{z}}-2\bar{Q}\partial_{\bar{Q}}-U\partial_U,\quad e_6=\bar{z}^2\partial_{\bar{z}}-4\bar{z}\bar{Q}\partial_{\bar{Q}}-2\bar{z}U\partial_U,\\
T_i=\partial_{F_i},\qquad i=1,2,3\\
D_i=F_i\partial_{F_{(i)}}+N_i\partial_{N_{(i)}},\\
R_{ij}=(F_i\partial_{F_j}-F_j\partial_{F_i})+(N_i\partial_{N_j}-N_j\partial_{N_i}),\qquad i< j=2,3\\
S_{ij}=(F_i\partial_{F_j}+F_j\partial_{F_i})+(N_i\partial_{N_j}+N_j\partial_{N_i}),
\ea
\ee
with nonzero commutation relations
\be
\ba{l}\hspace{-0.2cm}\ba{lll}
[e_1,e_3]=e_1,& [e_1,e_5]=-2e_3,& [e_3,e_5]=e_5,\\{}
[e_2,e_4]=e_2,& [e_2,e_6]=-2e_4,& [e_4,e_6]=e_6,\\{}
[e_0,T_i]=-T_i,& [T_i,D_j]=\delta_{ij}T_i,& [T_i,R_{jk}]=\delta_{ij}T_k-\delta_{ik}T_j,\ea\\{}
[T_i,S_{jk}]=\delta_{ij}T_k+\delta_{ik}T_j,\\{}
[D_i,R_{jk}]=\delta_{ij}S_{ik}-\delta_{ik}S_{ij},\qquad [D_i,S_{jk}]=\delta_{ij}R_{ik}-\delta_{ik}R_{ji},\\{} [R_{ij},S_{kl}]=\delta_{jk}S_{il}+\delta_{jl}S_{ik}-\delta_{ik}S_{jl}-\delta_{il}S_{jk}.
\ea
\ee
On the other hand the maximal finite-dimensional Lie algebra $A$ of the GC equations (\ref{claGC}) is spanned by the seven generators \cite{BGH}
\be\hspace{-1cm}\ba{l}
e_0=-H\partial_H+Q\partial_Q+\bar{Q}\partial_{\bar{Q}}+2U\partial_U,\\
e_1=\partial_z,\quad e_3=z\partial_z-2Q\partial_Q-U\partial_U,\quad e_5=z^2\partial_z-4zQ\partial_Q-2zU\partial_U,\\
e_2=\partial_{\bar{z}},\quad e_4=\bar{z}\partial_{\bar{z}}-2\bar{Q}\partial_{\bar{Q}}-U\partial_U,\quad e_6=\bar{z}^2\partial_{\bar{z}}-4\bar{z}\bar{Q}\partial_{\bar{Q}}-2\bar{z}U\partial_U,
\ea\label{clagen}\ee
with nonzero commutation relations
\be\ba{l}
[e_1,e_3]=e_1,\qquad [e_1,e_5]=-2e_3,\qquad [e_3,e_5]=e_5,\\{}
[e_2,e_4]=e_2,\qquad [e_2,e_6]=-2e_4,\qquad [e_4,e_6]=e_6.
\ea\ee
This 7-dimensional Lie subalgebra $A$ can be decomposed as a direct sum of two simple subalgebras together with a one-dimensional algebra generated by $e_0$ :
\be
A=\lbrace e_1,e_3,e_5\rbrace\oplus\lbrace e_2,e_4,e_6\rbrace\oplus\lbrace e_0\rbrace.
\ee
Therefore the classification of the subalgebras of $A$ into conjugacy classes involves two copies of a 3-dimensional Lie algebra together with the center $\lbrace e_0\rbrace$. This classification was performed in \cite{BGH}.

The conjecture on an integrable class of differential equations in the sense of the soliton theory proposed in \cite{CGS},\cite{LST}, states that the set of Lie point symmetries of the non-parametric linear problem is a subset of the set of Lie point symmetries of the original system. If the original system is non-integrable (in the sense of soliton theory), then the sets of symmetries are equal. If the original system is integrable, then there exists at least one symmetry of the original system which is not a symmetry of the non-parametric linear problem. These additional symmetries can be used to introduce a spectral parameter into the linear problem which cannot be eliminated by a gauge transformation. Note that the insertion condition is necessary for one to introduce a spectral parameter in the linear system. In order to compare the Lie point symmetries of the original system and those of the non-parametric linear problem, in the case where both sets of symmetries are finite-dimensional, we define the differential projector operator $\pi$ in the form of a dilation operator involving all independent and dependent variables
\be
\pi(L)=L(z\partial+\bar{z}\bar{\partial}+H\partial_H+Q\partial_Q+\bar{Q}\partial_{\bar{Q}}+U\partial_U),
\ee
where all the elements of the algebra $L$ are applied on the operator $\omega=z\partial+\bar{z}\bar{\partial}+H\partial_H+Q\partial_Q+\bar{Q}\partial_{\bar{Q}}+U\partial_U$. Note that $\omega$ is not necessarily an element of the algebra $L$ under consideration. This projector $\pi$ has the property that $\pi^n(L)=\pi(L)$ for any finite integer $n\in\mathbb{Z}^+$. In view of the above conjecture, we can characterize the integrability of the original system as follows. We denote by $L_1$ and $L_2$ the sets of Lie point symmetries of the original system and its non-parametric linear problem, respectively. If $L_1=\pi(L_2)$, then the original system is non-integrable (in the sense of the soliton theory). If there exist reductions of the original system (whose set of symmetries is $L_1'$) and of the non-parametric linear problem (whose set of symmetries is $L_2'$) such that $L_1'\neq\pi(L_2')$, then the reduced subsystem of the original system can be integrable. In view of the above statements, we observe that the ZCC for the GC equations (\ref{claGC}) is not an integrable system, since the algebra $\pi(L)$ and $A$ are equal
\be
A=\pi(L)=\lbrace e_1,e_3,e_5\rbrace\oplus\lbrace e_2,e_4,e_6\rbrace\oplus\lbrace e_0\rbrace.
\ee

It seems worthwhile to try to extend the conjecture proposed for the classical system to the case of a SUSY version of the GC equations and their reduced SUSY systems. This conjecture can be illustrated with the example of the SUSY sine-Gordon equation \cite{Siddiq05}
\be
D_+D_-\Phi=i\sin\Phi,\qquad D_\pm=\partial_{\theta^\pm}-i\theta^\pm\partial x_\pm,\label{sG}
\ee
where $\Phi$ is a real bosonic superfield and $D_\pm$ are covariant derivatives. Here, $x_+$ and $x_-$ are even Grassmann variables which constitute the orthogonal light-cone coordinates $x_\pm=\frac{1}{2}(t\pm x)$, while $\theta^+$ and $\theta^-$ are odd Grassmann variables. The symmetries which leave the SUSY equation (\ref{sG}) invariant are generated by the infinitesimal vector fields
\be
\ba{ll}
P_+=\partial_{x_+}, & \hspace{-1.5cm}P_-=\partial_{x_-},\\
J_+=\partial_{\theta^+}+i\theta^+\partial_{x_+}, & \hspace{-1.5cm}J_-=\partial_{\theta^-}+i\theta^-\partial_{x_-},\\
R=2x_+\partial_{x_+}+\theta^+\partial_{\theta^+}-2x_-\partial_{x_-}-\theta^-\partial_{\theta^-}. 
\ea
\ee
The non-parametric linear system corresponding to the SUSY sine-Gordon equation (\ref{sG}) is given by
\be
D_+\Psi=A_+\Psi,\qquad D_-\Psi=A_-\Psi,\label{sGLP}
\ee
where
\be
\hspace{-2.5cm}A_+=\frac{1}{2}\left(\ba{ccc}
0 & 0 & ie^{i\Phi} \\
0 & 0 & -ie^{-i\Phi} \\
-e^{-i\Phi} & e^{i\Phi} & 0
\ea\right),\qquad A_-=\left(\ba{ccc}
iD_-\Phi & 0 & -i \\
0 & -iD_-\Phi & i \\
-1 & 1 & 0
\ea\right).
\ee
The mixed derivatives of $\Psi$ are
\be\hspace{-2.5cm}\ba{l}
D_-D_+\Psi=D_-(A_+\Psi)=(D_-A_+)\Psi-EA_+E(D_-\Psi)=(D_-A_+)\Psi-EA_+EA_-\Psi,\\
D_+D_-\Psi=D_+(A_-\Psi)=(D_+A_-)\Psi-EA_-E(D_+\Psi)=(D_+A_-)\Psi-EA_-EA_+\Psi,
\ea\ee
where
\be
E=\pm\left(\ba{ccc}
1 & 0 & 0 \\
0 & 1 & 0 \\
0 & 0 &-1
\ea\right).\label{E}
\ee
Hence, the ZCC of equation (\ref{sGLP}) is given by
\be
D_-A_++D_+A_--\lbrace EA_+,EA_-\rbrace=0,\label{ZCCsG}
\ee
whenever equation (\ref{sG}) holds. The brackets $\lbrace\cdot,\cdot\rbrace$ denote the anticommutator, unless otherwise indicated. The ZCC (\ref{ZCCsG}) is equivalent to the SUSY sine-Gordon equation (\ref{sG}). The non-parametric linear system (\ref{sGLP}) is invariant under the transformation associated with the vector fields $P_\pm$ and $J_\pm$ but is not invariant with respect to the dilation $R$. This fact allows us to introduce a one parameter group associated with the dilation $R$ through the transformation $\tilde{x}_+=e^{2\mu}x_+$, $\tilde{x}_-=e^{-2\mu}x_-$, $\tilde{\theta}^+=e^\mu\theta^+$ and $\tilde{\theta}^-=e^{-\mu}\theta^-$, $\mu\in\mathbb{R}$, into the linear system (\ref{sGLP}) which gives us
\be
D_+\Psi=A_+\Psi,\qquad D_-\Psi=A_-\Psi,
\ee
where
\be
\hspace{-2.5cm}A_+=\frac{1}{2\sqrt{\lambda}}\left(\ba{ccc}
 0 & 0 & ie^{i\Phi} \\
 0 & 0 & -ie^{-i\Phi} \\
 -e^{-i\Phi} & e^{i\Phi} & 0 
\ea\right),~ A_-=\sqrt{\lambda}\left(\ba{ccc}
 \frac{i}{\sqrt{\lambda}}D_-\Phi & 0 & -i \\
 0 & -\frac{i}{\sqrt{\lambda}}D_-\Phi & i \\
 -1 & 1 & 0 
\ea\right).
\ee
This result coincides with the one established for the SUSY Lax pair found in \cite{Siddiq05}. Here, $\lambda=e^{2\mu}$ plays the role of a spectral parameter. The connection between the super-Darboux transformations and the super-B\"acklund transformations allows the construction of $n$ explicit supersoliton solutions \cite{Siddiq06}. The results obtained for the SUSY sine-Gordon equations (\ref{sG}) were so promising, that it seemed to be wothwhile to try to extend this approach and check its effectiveness for the case of a fermionic SUSY extension of the GC equations.

\section{Fermionic supersymmetric extension of the Gauss-Weingarten and Gauss-Codazzi equations} \setcounter{equation}{0}
The purpose of this section is to establish a SUSY version of the GW and GC equations using a fermionic superfield representation of a surface in the superspace $\mathbb{R}^{(1,1\vert2)}$.
Consider a SUSY version of the differential equations which define surfaces in two-dimensional Minkowski space with the bosonic coordinates $x_+$ and $x_-$, and the fermionic (anti-commuting) variables $\theta^+$ and $\theta^-$. Let $\mathcal{S}$ be a smooth simply connected surface in the superspace $\mathbb{R}^{(1,1\vert2)}=\lbrace(x_+,x_-,\theta^+,\theta^-)\rbrace$ which we assume is conformally parametrized in the sense that the surface $\mathcal{S}$ is given by a vector-valued fermionic superfield $F(x_+,x_-,\theta^+,\theta^-)$ satisfying the normalization conditions of function $F$ (\ref{gij}) specified below. Such a fermionic superfield can be decomposed as
\be
\hspace{-2.5cm}F(x_+,x_-,\theta^+,\theta^-)=F_l(x_+,x_-)+\theta^+\varphi_l(x_+,x_-)+\theta^-\psi_l(x_+,x_-)+\theta^+\theta^-G_l(x_+,x_-),
\ee
where $F_l$ and $G_l$ are odd-valued fields, while $\varphi_l$ and $\psi_l$ are even-valued fields (for $l=1,2,3$). In what follows we use the same notation for the fermionic immersion function $F$ as in the classical case. Also let $D_\pm$ be two covariant derivatives given by
\be
D_\pm=\partial_{\theta^\pm}-i\theta^\pm\partial_{x_\pm}.
\ee
The covariant derivatives $D_+$ and $D_-$ have the property that they anticommute with the differential supersymmetry operators
\begin{equation}
J_+=\partial_{\theta^+}+i\theta^+\partial_{x_+},\qquad J_-=\partial_{\theta^-}+i\theta^-\partial_{x_-},\label{J}
\end{equation}
which generate the SUSY transformations
\begin{equation}
x\rightarrow x+i\underline{\eta}_1\theta^+,\qquad \theta^+\rightarrow\theta^++i\underline{\eta}_1,\label{transformation0}
\end{equation}
and
\begin{equation}
t\rightarrow t+i\underline{\eta_2}\theta^-,\qquad \theta^-\rightarrow\theta^-+i\underline{\eta}_2,\label{transformation}
\end{equation}
respectively. Here $\underline{\eta}_1$ and $\underline{\eta}_2$ are odd-valued parameters. The four operators, $D_+$, $D_-$, $J_+$ and $J_-$ satisfy the following anticommutation relations
\begin{equation}
\hspace{-2.5cm}\lbrace J_m,J_n\rbrace=2i\delta_{mn}\partial_{x_m},\quad \lbrace D_m,D_n\rbrace=-2i\delta_{mn}\partial_{x_m},\quad \lbrace J_m,D_n\rbrace=0,\quad m,n=1,2\label{anticommutation}
\end{equation}
where $\delta_{mn}$ is the Kronecker delta function. Therefore we have the following relations
\begin{equation}
D_\pm^2=-i\partial_\pm,\qquad J_\pm^2=i\partial_\pm.\label{relations}
\end{equation}
The conformal parametrization of the surface in the superspace $\mathbb{R}^{(1,1\vert2)}$ is assumed to give the following normalization of $F$
\be
\langle D_iF,D_jF\rangle=g_{ij}f,\qquad i,j=1,2\label{normalizationF}
\ee
where the indices $1,2$ stand for $+,-$ respectively and $f$ is a bosonic function of $x_+$ and $x_-$. The bosonic metric functions $g_{ij}$ are given by
\be
g_{11}=g_{22}=0,\qquad g_{12}=g_{21}=\frac{1}{2}e^\phi,\label{gij}
\ee
which represent the metric coefficients and are symmetric, in contrast to the case in \cite{BGH}. We have introduced the function $f$ which allows for the possibility that the inner product (\ref{normalizationF}) may be bodiless. This in turn allows us to consider larger types of surfaces.

Let the bosonic function $\phi$ be decomposed through an expansion in the fermionic variables $\theta^+$ and $\theta^-$ as
\be
\phi=\phi_0+\theta^+\phi_1+\theta^-\phi_2+\theta^+\theta^-\phi_3,
\ee
where $\phi_0$ and $\phi_3$ are bosonic functions and $\phi_1$ and $\phi_2$ are fermionic functions. The exponential of $\phi$ is given by
\be\ba{c}
e^\phi=e^{\phi_0}(1+\theta^+\phi_1+\theta^-\phi_2+\theta^+\theta^-(\phi_3-\phi_1\phi_2)),\\
e^{-\phi}=e^{-\phi_0}(1-\theta^+\phi_1-\theta^-\phi_2-\theta^+\theta^-(\phi_3+\phi_1\phi_2)),
\ea\ee
The tangent vectors $D_+F$ and $D_-F$, together with the normal bosonic superfield $N$, form a bosonic moving frame $\Omega$ on the surface in the superspace $\mathbb{R}^{(1,1\vert2)}$. The normal vector $N$ can be decomposed as 
\be
\hspace{-2.5cm}N=N_m(x_+,x_-)\hspace{-0.1cm}+\hspace{-0.1cm}\theta^+\alpha_m(x_+,x_-)\hspace{-0.1cm}+\hspace{-0.1cm} \theta^-\beta_m(x_+,x_-)\hspace{-0.1cm}+\hspace{-0.1cm}\theta^+\theta^-H_m(x_+,x_-),\quad m=1,2,3
\ee
where $N_m$ and $H_m$ are even-valued fields and $\alpha_m$ and $\beta_m$ are odd-valued fields and have to satisfy the normalization
\be
\langle D_iF,N\rangle=0,\qquad \langle N,N\rangle=1,\qquad i=1,2.\label{normalizationN}
\ee

We assume that the second-order covariant derivatives of $F$ and the first-order covariant derivatives of $N$ are spanned by $D_\pm F$ and $N$ in such a way that
\be
D_jD_iF=\Gamma_{ij}^{\phantom{ij}k}D_kF+b_{ij}fN,\qquad D_iN=b^k_{\phantom{k}i}D_kF+\omega_iN,\label{decomp}
\ee
where $\Gamma_{ij}^{\phantom{ij}k}$, $b_{ij}$, $b^k_{\phantom{k}i}$ and $\omega_i$ are all fermionic functions.

\paragraph{}The GW equations for the moving frame $\Omega$ on a surface are given by
\be
D_+\Omega=A_+\Omega,\qquad D_-\Omega=A_-\Omega,\qquad \Omega=\left(\ba{c}D_+F\\D_-F\\N\ea\right),
\ee
where the $3\times3$ fermionic-valued matrices $A_+$ and $A_-$ are
\be
A_+=\left(\ba{ccc}
\Gamma_{11}^{\phantom{11}1} & \Gamma_{11}^{\phantom{11}2} & b_{11}f \\
\Gamma_{21}^{\phantom{21}1} & \Gamma_{21}^{\phantom{21}2} & b_{21}f \\
b^1_{\phantom{1}1} & b^2_{\phantom{2}1} & \omega_1
\ea\right),\qquad A_-=\left(\ba{ccc}
\Gamma_{12}^{\phantom{12}1} & \Gamma_{12}^{\phantom{12}2} & b_{12}f \\
\Gamma_{22}^{\phantom{22}1} & \Gamma_{22}^{\phantom{22}2} & b_{22}f \\
b^1_{\phantom{1}2} & b^2_{\phantom{2}2} & \omega_2
\ea\right).
\ee
We can obtain the conditions on $b_{ij}$, $b^i_{\phantom{i}j}$ and $\omega_i$ from the derivatives of (\ref{normalizationN})
\be\hspace{-2.5cm}\ba{l}
0=D_i\langle N,N\rangle=\langle D_iN,N\rangle+\langle N,D_iN\rangle=2\langle D_iN,N\rangle=2\omega_i\langle N,N\rangle,\\
0=D_i\langle D_jF,N\rangle=\langle D_iD_jF,N\rangle+\langle D_jF,D_iN\rangle=b_{ji}f\langle N,N\rangle+b^k_{\phantom{k}i}\langle D_jF,D_kF\rangle,
\ea\ee
which imply
\be
\omega_i=0,\qquad f(b_{ji}+g_{jk}b^k_{\phantom{k}i})=0,\qquad i,j=1,2.
\ee
The conditions on $\Gamma_{ij}^{\phantom{ij}k}$ can be obtained by taking the derivatives of equations (\ref{normalizationF}). We get
\be
\hspace{-2.5cm}0\hspace{-0.1cm}=\hspace{-0.1cm}D_i\langle D_jF,D_jF\rangle\hspace{-1mm}=\hspace{-1mm}\langle D_iD_jF,D_jF\rangle+\langle D_jF,D_iD_jF\rangle\hspace{-1mm}=\hspace{-1mm}2\langle D_iD_jF,D_jF\rangle\hspace{-1mm}=\hspace{-1mm}2\Gamma_{ji}^{\phantom{ji}k}g_{kj}.
\ee
Therefore we have
\be
\Gamma_{ji}^{\phantom{ji}^k}=0,\qquad\mbox{for}\qquad j\neq k
\ee
and by construction we have that $\Gamma_{ji}^{\phantom{ji}^k}$ is antisymmetric under a permutation of $i$ and $j$, i.e.
\be
\Gamma_{ji}^{\phantom{ji}^k}=-\Gamma_{ji}^{\phantom{ij}^k},\qquad\mbox{for}\qquad i\neq j.
\ee
This implies that
\be
\Gamma_{ji}^{\phantom{ji}^k}=0,\qquad\mbox{if}\qquad i\neq k\mbox{ or }j\neq k.\label{Gamma0}
\ee
Differentiating (\ref{normalizationF}) by the covariant derivatives $D_i$ we get
\be
\hspace{-2.5cm}\ba{l}\frac{1}{2}D_i(e^\phi f)=D_i\langle D_+F,D_-F\rangle=\langle D_iD_+F,D_-F\rangle+\langle D_+F,D_iD_-F\rangle\\
\phantom{\frac{1}{2}D_i(e^\phi f)}=\Gamma_{1i}^{\phantom{1i}1}\langle D_+F,D_-F\rangle+\Gamma_{2i}^{\phantom{2i}2}\langle D_+F,D_-F\rangle,\\
\frac{1}{2}e^\phi fD_i\phi+\frac{1}{2}e^\phi D_if=\frac{1}{2}e^\phi\Gamma_{1i}^{\phantom{1i}1}f+\frac{1}{2}e^\phi\Gamma_{2i}^{\phantom{2i}2}f,
\ea
\ee
which implies
\be
D_if=(\Gamma_{1i}^{\phantom{1i}1}+\Gamma_{2i}^{\phantom{2i}2}-D_i\phi)f.
\ee
However, using equation (\ref{Gamma0}), we get
\be
D_if=(\Gamma_{i(i)}^{\phantom{i(i)}(i)}-D_i\phi)f,\qquad i=1,2\label{Df}
\ee
where we do not sum over the index $i$. It is interesting to note that if $D_if=0$, then we have
\be
\Gamma_{11}^{\phantom{11}1}=D_+\phi,\qquad \Gamma_{22}^{\phantom{22}2}=D_-\phi,\label{Dfconstant}
\ee
which is similar to the classical case. Also, one can compute the compatibility condition on $f$ which is given by
\be
\lbrace D_+,D_-\rbrace f=(D_+\Gamma_{22}^{\phantom{22}2}+D_-\Gamma_{11}^{\phantom{11}1})f=0.\label{DDf}
\ee
The Christoffel symbols of the first kind are defined as
\be
\Gamma_{ijk}f=\langle D_jD_iF,D_kF\rangle,
\ee
and using equation (\ref{decomp}) we get the following relation between the Christoffel symbols of the first and second kinds
\be
\Gamma_{ijk}f=\langle D_jD_iF,D_kF\rangle=\Gamma_{ij}^{\phantom{ij}l}\langle D_lF,D_kF\rangle=\Gamma_{ij}^{\phantom{ij}l}g_{lk}f,
\ee
that is
\be
f(\Gamma_{ijk}-\Gamma_{ij}^{\phantom{ij}l}g_{lk})=0.
\ee
So we obtain
\be\ba{llll}
\Gamma_{111}=0, & \Gamma_{112}=\frac{1}{2}e^\phi \Gamma_{11}^{\phantom{11}1}, & \Gamma_{121}=0, & \Gamma_{211}=0,\\
\Gamma_{122}=0, & \Gamma_{212}=0, & \Gamma_{221}=\frac{1}{2}e^\phi \Gamma_{22}^{\phantom{22}2}, & \Gamma_{222}=0,
\ea\ee
up to the addition of a fermionic function $\zeta\neq0$ which has the property $\zeta f=0$. 

We define the fermionic quantities $b_{ij}$ to be
\be
b_{11}=Q^+,\qquad b_{12}=-b_{21}=\frac{1}{2}e^\phi H,\qquad b_{22}=Q^-,
\ee
which give the relations
\be
\hspace{-1.5cm}\langle D_+^2F,N\rangle=Q^+f,\qquad\langle D_-D_+F,N\rangle=\frac{1}{2}e^{\phi}H f,\qquad\langle D_-^2F,N\rangle=Q^-f.\label{Q+Q-H}
\ee
Hence, this implies that the fermionic quantities $b^i_{\phantom{i}j}$ are
\be
\hspace{-1cm}b^1_{\phantom{1}1}=H,\qquad b^2_{\phantom{2}1}=-2e^{-\phi}Q^+,\qquad b^1_{\phantom{1}2}=-2e^{-\phi}Q^-,\qquad b^2_{\phantom{2}2}=-H,
\ee
up to the addition of a fermionic function $\ell\neq0$ such that $\ell f=0$. The Gauss-Weingarten equations take the Bianchi form \cite{Bianchi}
\be\ba{c}
D_+\left(\ba{c}D_+F\\D_-F\\N\ea\right)=\left(\ba{ccc}
R^+ & 0 & Q^+f \\
0 & 0 & -\frac{1}{2}e^\phi Hf \\
H & -2e^{-\phi}Q^+ & 0
\ea\right)\left(\ba{c}D_+F\\D_-F\\N\ea\right),\\
\\
D_-\left(\ba{c}D_+F\\D_-F\\N\ea\right)=\left(\ba{ccc}
0 & 0 & \frac{1}{2}e^\phi Hf \\
0 & R^- & Q^-f \\
-2e^{-\phi}Q^- & -H & 0
\ea\right)\left(\ba{c}D_+F\\D_-F\\N\ea\right),
\ea\label{GW}\ee
where we define $R^+=\Gamma_{11}^{\phantom{11}1}$ and $R^-=\Gamma_{22}^{\phantom{22}2}$. The GC equations are the compatibility conditions of the GW equations. The ZCC for equations (\ref{GW}) in matrix form is
\be
D_+A_-+D_-A_+-\lbrace A_+,A_-\rbrace=0,\label{ZCC}
\ee
where
\be\ba{l}
A_+=\left(\ba{ccc}
\Gamma_{11}^{\phantom{11}1} & \Gamma_{11}^{\phantom{11}2} & b_{11}f \\
\Gamma_{21}^{\phantom{21}1} & \Gamma_{21}^{\phantom{21}2} & b_{21}f \\
b^1_{\phantom{1}1} & b^2_{\phantom{2}1} & \omega_1
\ea\right)=\left(\ba{ccc}
R^+ & 0 & Q^+f \\
0 & 0 & -\frac{1}{2}e^\phi Hf \\
H & -2e^{-\phi}Q^+ & 0
\ea\right),\\
 \\
A_-=\left(\ba{ccc}
\Gamma_{12}^{\phantom{12}1} & \Gamma_{12}^{\phantom{12}2} & b_{12}f \\
\Gamma_{22}^{\phantom{22}1} & \Gamma_{22}^{\phantom{22}2} & b_{22}f \\
b^1_{\phantom{1}2} & b^2_{\phantom{2}2} & \omega_2
\ea\right)=\left(\ba{ccc}
0 & 0 & \frac{1}{2}e^\phi Hf \\
0 & R^- & Q^-f \\
-2e^{-\phi}Q^- & -H & 0
\ea\right).
\ea\label{A+A-}\ee
In component form the equations are
\be\ba{rl}
(i) & D_-R^++2e^{-\phi}Q^+Q^-f=0,\\
(ii) & \left[D_-Q^++\frac{1}{2}e^\phi D_+H+Q^+(D_-\phi-R^-)\right]f=0,\\
(iii) & D_+R^-+2e^{-\phi}Q^-Q^+f=0,\\
(iv) & \left[D_+Q^--\frac{1}{2}e^\phi D_-H+Q^-(D_+\phi-R^+)\right]f=0,\\
(v) & D_+Q^--\frac{1}{2}e^\phi D_-H+Q^-(D_+\phi-R^+)=0,\\
(vi) & D_-Q^++\frac{1}{2}e^\phi D_+H+Q^+(D_-\phi-R^-)=0.
\ea\label{preGC}\ee
If the equations (\ref{preGC}.vi) and (\ref{preGC}.v) hold, then equations (\ref{preGC}.ii) and (\ref{preGC}.iv) are identically satisfied respectively. By adding (\ref{preGC}.i) and (\ref{preGC}.iii) we obtain
\be
D_-R^++D_+R^-=0,
\ee
which is the compatibility condition for $f$, i.e. (\ref{DDf}). Therefore the Gauss-Codazzi equations are reduced to the four linearly independent equations
\be\ba{rl}
(i) & D_+R^-+D_-R^+=0,\\
(ii) & D_-R^++2e^{-\phi}Q^+Q^-f=0,\\
(iii) & D_+Q^--\frac{1}{2}e^\phi D_-H+Q^-(D_+\phi-R^+)=0,\\
(iv) & D_-Q^++\frac{1}{2}e^\phi D_+H+Q^+(D_-\phi-R^-)=0.
\ea\label{GC}\ee
The Grassmann-valued PDEs (\ref{GC}) involve six dependent functions of the independent variables $x_+$, $x_-$, $\theta^+$ and $\theta^-$ including one bosonic function $\phi$ and the five fermionic functions $H$, $Q^\pm$ and $R^\pm$ together with one bosonic function $f$ of $x_+$ and $x_-$. Hence, if we restrict ourselves to the case where $f$ is a bosonic constant, then from (\ref{Dfconstant}) the Gauss-Codazzi equations become
\be\ba{rl}
(i) & D_-D_+\phi+2e^{-\phi}Q^+Q^-f=0,\\
(ii) & D_+Q^--\frac{1}{2}e^\phi D_-H=0,\\
(iii) & D_-Q^++\frac{1}{2}e^\phi D_+H=0,
\ea\label{GC0}\ee
which resemble the classical GC equations (\ref{claGC}) taking into account that the $H^2$ term vanishes. The equations (\ref{GC0}) have some terms whose signs differ from those of the classical equations. We get an underdetermined system of three PDEs for four dependent variables $H$, $Q^\pm$ and $\phi$.

\paragraph{}Under the above assumptions we obtain the following result.

\begin{proposition}[Structural SUSY equations for a moving frame on a surface]
For any fermionic superfield $F(x_+,x_-,\theta^+,\theta^-)$ and bosonic superfield $N(x_+,x_-,\theta^+,\theta^-)$ satisfying the normalization conditions (\ref{normalizationF}), (\ref{gij}), (\ref{normalizationN}) and (\ref{Q+Q-H}), the bosonic moving frame $\Omega=(D_+F,D_-F,N)^T$ on a surface immersed in the superspace $\mathbb{R}^{(1,1\vert2)}$ satisfies the SUSY GW equations (\ref{GW}). The ZCC (\ref{ZCC}), which is the compatibility condition of the SUSY GW equations (\ref{GW}) expressed in terms of the matrices $A_+$ and $A_-$, is equivalent to the SUSY GC equations (\ref{GC}).
\end{proposition}

\section{Geometric aspects of the supersymetric fermionic conformally parametrized surfaces}\setcounter{equation}{0}
In this section, we discuss certain aspects of Grassmann variables in conjunction with differential geometry and supersymmetry analysis. Let us define $d_\pm=d\theta^\pm+idx_\pm\partial_{\theta^\pm}$ to be the infinitesimal displacement in the direction of $D_\pm$. The first fundamental form is given by
\be
\hspace{-1.5cm}\ba{l}
I=\left\langle\left(\ba{cc}d_+ & d_-\ea\right),\left(\ba{cc}d_+ & d_-\ea\right)\left(\ba{cc}
\langle D_+F,D_+F\rangle & \langle D_+F,D_-F\rangle\\
-\langle D_+F,D_-F\rangle & \langle D_-F,D_-F\rangle
\ea\right)\right\rangle\\
\phantom{I}=\left\langle\left(\ba{cc}d_+ & d_-\ea\right),\left(\ba{cc}d_+ & d_-\ea\right)\left(\ba{cc}
g_{11}f & g_{12}f\\
-g_{12}f & g_{22}f
\ea\right)\right\rangle\\
\phantom{I}=\left\langle\left(\ba{cc}d_+ & d_-\ea\right),\left(\ba{cc}d_+ & d_-\ea\right)Rf\right\rangle\\
\phantom{I}=f\left(d_+^2g_{11}+2d_+d_-g_{12}+d_-^2g_{22}\right)=fd_+d_-e^\phi,\label{I}
\ea
\ee
and the discriminant $g$ is defined to be
\be
g=g_{11}g_{22}+g_{12}^2=\frac{1}{4}e^{2\phi}.\label{discg}
\ee
Hence the covariant metric is given by
\be
g_{ij}g^{jk}=\delta_i^k,\qquad \left(\ba{cc}
g_{11} & g_{21} \\
g_{12} & g_{22}
\ea\right)\left(\ba{cc}
g^{11} & g^{21} \\
g^{12} & g^{22}
\ea\right)=\left(\ba{cc}
1 & 0 \\
0 & 1
\ea\right),
\ee
such that
\be
g^{11}=g^{22}=0,\qquad g^{12}=g^{21}=2e^{-\phi} .
\ee
The second fundamental form is
\be
\hspace{-1.5cm}\ba{l}
I\hspace{-0.1cm}I=\left\langle\left(\ba{cc}d_+ & d_-\ea\right),\left(\ba{cc}d_+ & d_-\ea\right)\left(\ba{cc}
\langle D_+^2F,N\rangle & \langle D_-D_+F,N\rangle\\
-\langle D_-D_+F,N\rangle & \langle D_-^2F,N\rangle
\ea\right)\right\rangle\\
\phantom{I}=\left\langle\left(\ba{cc}d_+ & d_-\ea\right),\left(\ba{cc}d_+ & d_-\ea\right)\left(\ba{cc}
b_{11}f & b_{12}f\\
-b_{12}f & b_{22}f
\ea\right)\right\rangle\\
\phantom{I}=\left\langle\left(\ba{cc}d_+ & d_-\ea\right),\left(\ba{cc}d_+ & d_-\ea\right)Sf\right\rangle\\
\phantom{I}=f\left(d_+^2b_{11}+2d_+d_-b_{12}+d_-^2b_{22}\right)=f\left(d_+^2Q^++d_+d_-(e^\phi H)+d_-^2Q^-\right),
\ea\label{II}
\ee
and the discriminant is defined to be
\be
b=b_{11}b_{22}+b_{12}^2=b_{11}b_{22}=Q^+Q^-.\label{discb}
\ee
In order to compute the first and second fundamental forms, we have assumed that $(d\theta^j\_\hspace{-0.15cm}\shortmid\partial_{\theta^i})=0$ for $i, j=1,2.$ Making use of (\ref{discg}) and (\ref{discb}), the Gaussian curvature is defined as
\be
\mathcal{K}=\det(SR^{-1})=\frac{b_{11}b_{22}+(b_{12})^2}{g_{11}g_{22}+(g_{12})^2}=4e^{-2\phi}Q^+Q^-,\label{Gaussian}
\ee
which is a bosonic bodiless function, where
\begin{equation*}
\hspace{-2.5cm}R=\left(\ba{cc}
g_{11} & g_{12}\\
-g_{12} & g_{22}
\ea\right)=\frac{1}{2}e^\phi\left(\ba{cc}
0 & 1\\
-1 & 0
\ea\right),\quad S=\left(\ba{cc}
b_{11} & b_{12}\\
-b_{12} & b_{22}
\ea\right)=\left(\ba{cc}
Q^+ & \frac{1}{2}e^\phi H\\
-\frac{1}{2}e^\phi H & Q^-
\ea\right),
\end{equation*}
while the mean curvature $H$ is a fermionic function satisfying
\be
H=\frac{1}{2}\mbox{tr}(SR^{-1}).
\ee

\paragraph{}Under the above assumptions on the SUSY version of the GC equations (\ref{GC}) we can provide a SUSY analogue of the Bonnet Theorem.
\begin{proposition}[Fermionic supersymmetric extension of the Bonnet theorem]
Given a SUSY conformal metric, $M=fd_+d_-e^\phi$, of a conformally parametrized surface $\mathcal{S}$, the Hopf differentials $d_\pm^2Q^\pm $ and a mean curvature function $H$ defined on a Riemann surface $\mathcal{R}$ satisfying the GC equation (\ref{GC}), there exists a vector-valued fermionic immersion function, $F=(F_1,F_2,F_3):\tilde{\mathcal{R}}\rightarrow\mathbb{R}^{(1,1\vert2)},$ with the fundamental forms
\begin{equation}
I= fd_+d_-e^\phi,\qquad I\hspace{-0.1cm}I=f(d_+^2Q^++ d_+d_-(He^\phi)+d_-^2Q^-),
\end{equation}
where $\tilde{\mathcal{R}}$ is the universal covering of the Riemann surface $\mathcal{R}$ and $\mathbb{R}^{(1,1\vert2)}$ is the superspace. The immersion function $F$ is unique up to affine transformations in the superspace $\mathbb{R}^{(1,1\vert2)}$.
\end{proposition}

\paragraph{}The proof of this proposition is analogous to that given in \cite{Bonnet}. Note that it is straightforward to construct surfaces on the superspace $\mathbb{R}^{(1,1\vert2)}$ related to integrable equations using a SUSY version of the Sym-Tafel formula for immersion \cite{Sym},\cite{Tafel}. However, it is nontrivial to identify those surfaces which have an invariant geometrical characterization. A list of such surfaces is known in the classical case \cite{Bob} but, to our knowledge, an identification of such surfaces is an open problem in the case of surfaces immersed in a superspace.

\section{Symmetries of the fermionic supersymmetric Gauss-Codazzi equations}\setcounter{equation}{0}
By a symmetry supergroup $G$ of a SUSY system, we mean a local supergroup of transformations acting on the Cartesian product $\mathcal{X}\times\mathcal{U}$ of supermanifolds, where $\mathcal{X}$ is the space of four independent variables $(x_+,x_-,\theta^+,\theta^-)$ and $\mathcal{U}$ is the space of seven dependent superfields $(\phi,H,Q^+,Q^-,R^+,R^-,f)$. Solutions of the GC equations (\ref{GC}) are mapped to solutions of (\ref{GC}) by the action of the group $G$ on the functions $\phi,H,Q^+,Q^-,R^+,R^-$ and $f$ of $(x_+,x_-,\theta^+,\theta^-)$. When we perform the symmetry reductions, we need to take into consideration the fact that the bosonic function $f$ introduced in (\ref{normalizationF}) depends only on $x_+$ and $x_-$ or is constant. If $G$ is a Lie supergroup as described in \cite{Kac} and \cite{Winternitz}, it can be associated with its Lie superalgebra $\mathfrak{g}$ whose elements are infinitesimal symmetries of equations (\ref{GC}). We have made use of the theory described in the book by Olver \cite{Olver} in order to determine a superalgebra of infinitesimal symmetries. 
The SUSY GC equations (\ref{GC}) are invariant under the following six bosonic symmetry generators
\renewcommand{\theequation}{\thesection.\arabic{equation}a}
\be
\ba{l}
P_+=\partial_{x_+},\qquad P_-=\partial_{x_-},\\

C_0=H\partial_H+Q^+\partial_{Q^+}+Q^-\partial_{Q^-}-2f\partial_f,\\
K_0=-H\partial_H+Q^+\partial_{Q^+}+Q^-\partial_{Q^-}+2\partial_\phi,\\
K_1=-2x_+\partial_{x_+}-\theta^+\partial_{\theta^+}+2Q^+\partial_{Q^+}+R^+\partial_{R^+}+\partial_\phi,\\
K_2=-2x_-\partial_{x_-}-\theta^-\partial_{\theta^-}+2Q^-\partial_{Q^-}+R^-\partial_{R^-}+\partial_\phi,
\ea\label{sym}
\ee
together with the three fermionic generators\setcounter{equation}{0}\renewcommand{\theequation}{\thesection.\arabic{equation}b}
\be
J_+=\partial_{\theta^+}+i\theta^+\partial_{x_+},\qquad J_-=\partial_{\theta^-}+i\theta^-\partial_{x_-},\qquad W=\partial_H.
\ee
\renewcommand{\theequation}{\thesection.\arabic{equation}}
The symmetry generators $W$ and $P_\pm$ represent a fermionic translation of $H$ and bosonic translations in the $x_\pm$ direction respectively, $J_\pm$ represent SUSY transformations and $C_0$, $K_0$, $K_1$ and $K_2$ represent dilations. The commutation table (anticommutation for two fermionic symmetries) for the generators of the superalgebra of equations  (\ref{GC}) is given in Table~1.
\begin{table}[h!]
\centering
\caption{Commutation table for the Lie superalgebra $\mathfrak{g}$ spanned by\\ the vector fields (6.1). In the case of two fermionic generators $J_+$\\ and/or $J_-$ and/or $W$ we have anticommutation rather than commutation.}
\begin{tabular}{c|c|c|c|c|c|c|c|c|c|}
 &$K_1$&$P_+$&$J_+$&$K_2$&$P_-$&$J_-$&$K_0$&$C_0$&$W$\\
\hline$K_1$&$0$ &$2P_+$ &$J_+$ &$0$ &$0$ &$0$ &$0$&$0$&$0$ \\
\hline$P_+$&$-2P_+$ &$0$ &$0$ &$0$ &$0$ &$0$ &$0$&$0$&$0$ \\
\hline$J_+$&$-J_+$ &$0$ &$2iP_+$ &$0$ &$0$ &$0$ &$0$&$0$&$0$ \\
\hline$K_2$&$0$ &$0$ &$0$ &$0$ &$2P_-$ &$J_-$ &$0$&$0$&$0$ \\
\hline$P_-$&$0$ &$0$ &$0$ &$-2P_-$ &$0$ &$0$ &$0$&$0$&$0$ \\
\hline$J_-$&$0$ &$0$ &$0$ &$-J_-$ &$0$  &$2iP_-$ &$0$&$0$&$0$ \\
\hline$K_0$&$0$ &$0$ &$0$ &$0$ &$0$ &$0$ &$0$&$0$&$W$ \\
\hline$C_0$&$0$ &$0$ &$0$ &$0$ &$0$ &$0$ &$0$&$0$&$-W$ \\
\hline$W$&$0$ &$0$ &$0$ &$0$ &$0$ &$0$ &$-W$&$W$ &$0$ \\
\hline
\end{tabular}
\end{table}

\noindent The decomposition of the superalgebra (\ref{GC}) is given by
\be
\hspace{-1cm}\mathfrak{g}=\lbrace\lbrace K_1\rbrace\sdir\lbrace P_+,J_+\rbrace\rbrace \oplus\lbrace\lbrace K_2\rbrace\sdir\lbrace P_-,J_-\rbrace\rbrace\oplus\lbrace\lbrace K_0,C_0\rbrace\sdir\lbrace W\rbrace\rbrace.\label{cla}
\ee
In equation (\ref{cla}) the braces $\lbrace\cdot,\cdot\rbrace$ denote the set of generators listed in (6.1).

However, if we consider the case where $D_\pm f=0$, the equations (\ref{GC0}) are invariant under the five bosonic generators\renewcommand{\theequation}{\thesection.\arabic{equation}a}
\be
\ba{l}
P_+=\partial_{x_+},\qquad P_-=\partial_{x_-},\\
K_0=-H\partial_H+Q^+\partial_{Q^+}+Q^-\partial_{Q^-}+2\partial_\phi,\\
K_1=-2x_+\partial_{x_+}-\theta^+\partial_{\theta^+}+2Q^+\partial_{Q^+}+\partial_\phi,\\
K_2=-2x_-\partial_{x_-}-\theta^-\partial_{\theta^-}+2Q^-\partial_{Q^-}+\partial_\phi,
\ea\label{sym0}
\ee
and the three fermionic generators\setcounter{equation}{2}\renewcommand{\theequation}{\thesection.\arabic{equation}b}
\be
J_+=\partial_{\theta^+}+i\theta^+\partial_{x_+},\qquad J_-=\partial_{\theta^-}+i\theta^-\partial_{x_-},\qquad W=\partial_H.
\ee
The commutation table for the generators of the superalgebra of equations  (\ref{GC0}) is given in Table 2.\renewcommand{\theequation}{\thesection.\arabic{equation}}
\begin{table}[h!]
\centering
\caption{Commutation table for the Lie superalgebra $\mathfrak{g}$ spanned by\\ the vector fields (6.3). In the case of two fermionic generators $J_+$\\ and/or $J_-$ and/or $W$ we have anticommutation rather than commutation.}
\begin{tabular}{c|c|c|c|c|c|c|c|c|}
 &$K_1$&$P_+$&$J_+$&$K_2$&$P_-$&$J_-$&$K_0$&$W$\\
\hline$K_1$&$0$ &$2P_+$ &$J_+$ &$0$ &$0$ &$0$ &$0$&$0$ \\
\hline$P_+$&$-2P_+$ &$0$ &$0$ &$0$ &$0$ &$0$ &$0$&$0$ \\
\hline$J_+$&$-J_+$ &$0$ &$2iP_+$ &$0$ &$0$ &$0$ &$0$&$0$ \\
\hline$K_2$&$0$ &$0$ &$0$ &$0$ &$2P_-$ &$J_-$ &$0$&$0$ \\
\hline$P_-$&$0$ &$0$ &$0$ &$-2P_-$ &$0$ &$0$ &$0$&$0$ \\
\hline$J_-$&$0$ &$0$ &$0$ &$-J_-$ &$0$  &$2iP_-$ &$0$&$0$ \\
\hline$K_0$&$0$ &$0$ &$0$ &$0$ &$0$ &$0$ &$0$&$W$ \\
\hline$W$&$0$ &$0$ &$0$ &$0$ &$0$ &$0$ &$-W$& $0$ \\
\hline
\end{tabular}
\end{table}
This Lie superalgebra $\mathfrak{g}$ can be decomposed into the following combination of direct and semi-direct sums
\begin{equation}
\mathfrak{g}=\lbrace\lbrace K_1\rbrace\sdir\lbrace P_+,J_+\rbrace\rbrace \oplus\lbrace\lbrace K_2\rbrace\sdir\lbrace P_-,J_-\rbrace\rbrace\oplus\lbrace\lbrace K_0\rbrace\sdir\lbrace W\rbrace\rbrace.\label{cla0}
\end{equation}
In equation (\ref{cla0}) the braces $\lbrace\cdot,...,\cdot\rbrace$ denote the set of generators listed in (6.3). The one-dimensional subalgebras of this superalgebra can be classified into conjugacy classes.

\section{One-dimensional subalgebras of the symmetry superalgebra $\mathfrak{g}$}\setcounter{equation}{0}
When constructing a list of representative one-dimensional subalgebras of the superalgebra $\mathfrak{g}$, it would be inadequate to consider the $\mathbb{R}$ or $\mathbb{C}$ span of the generators (6.1) because the odd generators $J_+,J_-$ and $W$ are multiplied by odd parameters in the list of subalgebras presented in Table~3. One is therefore led to consider a $\mathfrak{g}$ which is a supermanifold. That is, $\mathfrak{g}$ contains sums of any even combinations of $P^+,P^-,C_0,K_0,K_1$ and $K_2$ (multiplied by even parameters in $\Lambda_{even}$) and odd combinations of $J_+,J_-$ and $W$ (multiplied by odd parameters in $\Lambda_{odd}$). Therefore $\mathfrak{g}$ is an even Lie module. This leads to the following consideration. For a given $X\in\mathfrak{g}$, the subalgebras $\mathfrak{X}$ and $\mathfrak{X}'$ spanned by $X$ and $X'=aX$ with $a\in\Lambda_{even}\backslash\mathbb{C}$ are not isomorphic in general, i.e. $\mathfrak{X}'\subset\mathfrak{X}$.

It should be noted that subalgebras obtained by multiplying other subalgebras by bodiless elements of $\Lambda_{even}$ do not provide us with anything new for the purpose of symmetry reduction. It is not particularly useful to consider a subalgebra of the form e.g. $\lbrace P_++\underline{\eta}_1\underline{\eta}_2P_-\rbrace$, since there is no limit to the number of odd parameters $\underline{\eta}_k$ that can be used to construct even coefficients. While such subalgebras may allow for more freedom in the choice of invariants, we then encounter the problem of non-standard invariants \cite{BGH},\cite{GHS09},\cite{GHS11}. Such non-standard invariants, which do not lead to standard reductions or invariant solutions, are found for several other SUSY hydrodynamic-type systems \cite{GH11},\cite{GH13}. For the SUSY GC equations (\ref{GC}), the one-dimensional subalgebras $\mathfrak{g}_3,\mathfrak{g}_7,\mathfrak{g}_{19},\mathfrak{g}_{25},\mathfrak{g}_{46},\mathfrak{g}_{74}$ and $\mathfrak{g}_{158}$ listed in Table~3 have such non-standard invariants. 

Subgroups within the same conjugacy class lead to invariant solutions that are equivalent in the sense that a suitable symmetry can transform one to the other. It is therefore unnecessary to consider reductions with respect to algebras which are conjugate to each other. We make use of the methods for classifying direct and semi-direct sums of algebras as described in \cite{Winternitz} in order to classify the Lie superalgebra (\ref{cla}) under the action of the supergroup generated by $\mathfrak{g}$. More specifically, we generalize these methods to the case of a superalgebra involving both even and odd generators. Specifically, the Goursat twist method is used for the case of direct sums of algebras. Here the superalgebra (\ref{cla}) contains two isomorphic copies of the 3-dimensional algebra $\mathfrak{g}_1=\lbrace\lbrace K_1\rbrace\sdir\lbrace P_+,J_+\rbrace\rbrace$, the other copy being $\mathfrak{g}_2=\lbrace\lbrace K_2\rbrace\sdir\lbrace P_-,J_-\rbrace\rbrace$ together with the three-dimensional algebra $\lbrace\lbrace K_0,C_0\rbrace\sdir\lbrace W\rbrace\rbrace$ . This allows us to adapt the classification for $3$-dimensional algebras as described in \cite{Patera} to the SUSY case. We therefore begin our classification by considering the twisted one-dimensional subalgebras of $\mathfrak{g}_1\oplus\mathfrak{g}_2$. Under the action of a one-parameter group generated by the vector field
\be
X=\alpha K_1+\beta P_++\underline{\eta}J_++\delta K_2+\lambda P_-+\underline{\rho}J_-,
\ee
where $\alpha,\beta,\delta,\lambda\in\Lambda_{even}$ and $\underline{\eta},\underline{\rho}\in\Lambda_{odd}$, the one-dimensional subalgebra $Y=P_++aP_-,a\in\Lambda_{even}$ transforms under the Baker-Campbell-Hausdorff formula
\be
e^XYe^{-X}=Y+[X,Y]+\frac{1}{2!}[X,[X,Y]]+\frac{1}{3!}[X,[X,[X,Y]]]+...\label{BCHform}
\ee
to  $e^{-2\alpha}P_++e^{-2\delta}aP_-$. By an appropriate choice of $\alpha$ and $\delta$, the factor $e^{2\alpha-2\delta}a$ can be rescaled to either $1$ or $-1$. Hence, we get a twisted subalgebra $\mathfrak{g}_{14}=\lbrace P_++\epsilon P_-,\epsilon=\pm1\rbrace$.

As another example, consider a twisted algebra of the form $\lbrace K_1+\underline{\zeta}W\rbrace$, where $\underline{\zeta}$ is a fermionic parameter. Through the Baker-Campbell-Hausdorff formula (\ref{BCHform}), the vector field $Y=K_1+\underline{\zeta}W$ transforms through
\be
X=\alpha K_1+\beta P_++\underline{\eta}J_++\gamma K_2+\delta P_-+\underline{\lambda}J_-+\rho K_0+\sigma C_0+\underline{\tau}W,
\ee
(where $\alpha,\beta,\gamma,\delta,\rho,\sigma\in\Lambda_{even}$ and $\underline{\eta},\underline{\lambda},\underline{\tau}\in\Lambda_{odd}$) to 
\be
e^XYe^{-X}=K_1+e^{\rho-\sigma}\underline{\zeta}W-\frac{\beta}{\alpha}(e^{2\alpha}-1)P_+-\frac{1}{\alpha}(e^\alpha-1)\underline{\eta}J_+.\label{74}
\ee
Through a suitable choice of $\beta$ and $\underline{\eta}$, the last two terms of the expression (\ref{74}) can be eliminated, so we obtain the twisted subalgebra $\mathfrak{g}_{32}=\lbrace K_1+\underline{\zeta}W\rbrace$. Continuing the classification in a similar way, involving twisted and non-twisted subalgebras according to \cite{Winternitz}, we obtain the list of one-dimensional subalgebras given in Table~3 in the Appendix. These representative subalgebras allow us to determine invariant solutions of the SUSY GC equations (\ref{GC}) using the SRM. For the specific case where $f$ is constant (i.e. the SUSY GC equations (\ref{GC0})), the one-dimensional subalgebras of the resulting Lie symmetry superalgebra (\ref{cla0}) can be found by taking the limit where the coefficients of $C_0$ tend to zero in the subalgebra listed in Table~3 and withdrawing repeated subalgebras, while rescaling appropiately.

\section{Invariant solutions of the fermionic supersymmetric GC equations}\setcounter{equation}{0}
The SRM allows us to obtain invariant solutions of the GC equations (\ref{GC}). We proceed as follows. For each one-dimensional subalgebra listed in Table~3 (admitting standard invariants) we can find the orbit of the corresponding SUSY subgroup, which can be parametrized in terms of a bosonic symmetry variable $\xi$ and two fermionic symmetry variables, say $\eta$ and $\sigma$, which in turn are expressed in terms of $\theta^+$ and $\theta^-$, respectively. The superfields $\mathcal{U}=(H,Q^+,Q^-,R^+,R^-,\phi,f)$ are expanded in terms of the fermionic invariants $\eta$ and $\sigma$ with some coefficients expressed in terms of a bosonic symmetry variable $\xi$. Substituting these expanded forms of the superfields $\mathcal{U}$ into the GC equations (\ref{GC}) we reduce these equations to many possible differential subsystems involving even and odd functions. Solving these subsystems, we determine the invariant solutions and provide some geometrical interpretation of the associated surfaces. To illustrate this approach, we present three examples.

\paragraph{}\textbf{1.} For the subalgebra $\mathfrak{g}_{124}=\lbrace P_++\epsilon P_-+aK_0,\epsilon=\pm1,a\neq0\rbrace$, the orbit of the corresponding group of the SUSY GC equations (\ref{GC}) can be parametrized as follows
\be
\hspace{-2.5cm}\ba{ll}
H=e^{-ax_+}[h_0(\xi)+\theta^+\theta^- h_1(\xi)], & R^+=r^+_0(\xi)+\theta^+\theta^-r^+_1(\xi), \\
Q^+=e^{ax_+}[q_0^+(\xi)+\theta^+\theta^-q_1^+(\xi)], & R^-=r^-_0(\xi)+\theta^+\theta^-r^-_1(\xi), \\
Q^-=e^{ax_+}[q_0^-(\xi)+\theta^+\theta^-q_1^-(\xi)], & \phi=\varphi_0(\xi)+\theta^+\theta^-\varphi_1(\xi)+2ax_+,\quad f=\psi(\xi),
\ea
\ee
where the fermionic functions $H,Q^\pm$ and $R^\pm$ are expressed in terms of the bosonic symmetry variable $\xi=x_+-\epsilon x_-$ and the fermionic symmetry variables $\theta^+$ and $\theta^-$, while the bosonic functions $\varphi_0,\varphi_1$ and $\psi$ are expressed in terms of $\xi$ only. A corresponding invariant solution is given by
\be
\hspace{-2.5cm}\ba{l}
H=-2\underline{C}_0^+\underline{C}_0^-e^{-ax_+}\left[\epsilon e^{-\varphi_0}\underline{m}_0^++i\theta^+\theta^-(e^{-\varphi_0}\underline{m}_0^+)_\xi\right],\\
Q^+=-e^{ax_+}\underline{C}_0^+\underline{C}_0^-\left[ \underline{m}_0^++i\theta^+\theta^-((\underline{m}_0^+)_\xi+\epsilon a\underline{m}_0^+)\right],\\
Q^-=e^{ax_+}\underline{C}_0^+\underline{C}_0^-\left[\underline{m}_0^-+i\theta^+\theta^-(\epsilon a\underline{m}_0^-+(\underline{m}_0^-)_\xi)\right],\\
\phi=\varphi_0(\xi)+i\theta^+\theta^-(\varphi_0)_\xi+2ax_+,\quad 
R^+=\underline{C}_0^+,\quad R^-=\underline{C}_0^-,\quad f=\psi(\xi),
\ea\label{sol60}
\ee
where the fermionic functions $\underline{m}_0^+$, $\underline{m}_0^-$ and the bosonic function $\varphi_0$ of the symmetry variable $\xi$ satisfy the differential constraint
\begin{equation}
[e^{-\varphi_0}(\underline{m}_0^--\epsilon \underline{m}^+_0)]_\xi+\epsilon a\underline{m}_0^-e^{-\varphi_0}=0.
\end{equation}
Here $\psi$ is an arbitrary bosonic function of $\xi$, while $\underline{C}_0^+$ and $\underline{C}_0^-$ are arbitrary fermionic constants.

The first and second fundamental forms of the surface $S$ associated with the solution (\ref{sol60}) are given by
\be
\hspace{-2.5cm}\ba{l}
I=\psi e^{\varphi_0+2ax_+}d_+d_-\left[1+\theta^+\theta^-\varphi_1\right],\\
I\hspace{-0.1cm}I=e^{ax_+}\underline{C}_0^+\underline{C}_0^-\psi\left[-2d_+d_-\left(\epsilon \underline{m}_0^++i\theta^+\theta^-\left[(\underline{m}_0^+)_\xi-\epsilon i\varphi_1 \underline{m}_0^+-(\varphi_0)_\xi \underline{m}_0^+\right]\right)\right.\\
\hspace{7mm}\left.+d_+^2\left(\underline{m}_0^++i\theta^+\theta^-\left[(\underline{m}_0^+)_\xi+\epsilon a\underline{m}_0^+\right]\right)+d_-^2\left(\underline{m}_0^-+i\theta^+\theta^-\left[(\underline{m}_0^-)_\xi+\epsilon a\underline{m}_0^-\right]\right)\right].
\ea
\ee
The Gaussian curvature (\ref{Gaussian}) takes the form
\be
\mathcal{K}=0.
\ee

In particular when $a=0$, which corresponds to the subalgebra $\mathfrak{g}_{14}=\lbrace P_++\epsilon P_-\rbrace$, the orbits of the group of the SUSY GC equations (\ref{GC}) can be parametrized in such a way that $H, Q^\pm$ and $R^\pm$ are fermionic functions of the bosonic symmetry variable $\xi=x_--\epsilon x_+$, and the fermionic coordinates $\theta^+$ and $\theta^-$ while $\phi$ is a bosonic function of $\xi,\theta^+$ and $\theta^-$, and $\psi$ is a bosonic function of $\xi$ only. Under the assumption that the unknown functions take the form
\be
\ba{ll}
H=h_0(\xi)+\theta^+\theta^- h_1(\xi), & R^\pm=r_0^\pm(\xi)+\theta^+\theta^- r_1^\pm(\xi),\\
Q^\pm=q_0^\pm(\xi)+\theta^+\theta^-q_1^\pm(\xi), & \phi=\varphi_0(\xi)+\theta^+\theta^-\varphi_1(\xi), \qquad f=\psi(\xi),
\ea\label{sol61}
\ee
the corresponding invariant solution of the SUSY GC equations (\ref{GC}) is given by
\be
\hspace{-2.5cm}\ba{l}
H=2\underline{C}_0^-\underline{C}_0^+\underline{l}\left[\int e^{-\varphi_0}d\xi+i\theta^+\theta^-e^{-\varphi_0}\right]+\underline{C},\qquad \epsilon=1,\\
Q^+=\underline{C}_0^-\underline{C}_0^+\underline{l}e^{\varphi_0}\int e^{-\varphi_0}d\xi+\underline{C}_0^-B_0^+e^{\varphi_0}\\
\hspace{2.5cm}+i\theta^+\theta^-\underline{C}_0^-\left[\underline{C}_0^+\underline{l}\left(e^{\varphi_0}(\varphi_0)_\xi\int e^{-\varphi_0}d\xi+1\right)+B_0^+e^{\varphi_0}(\varphi_0)_\xi\right],\\
Q^-=\underline{C}_0^+\underline{C}_0^-\underline{l}e^{\varphi_0}\int e^{-\varphi_0}d\xi+\underline{C}_0^+B_0^-e^{\varphi_0}\\
\hspace{2.5cm}+i\theta^+\theta^-\underline{C}_0^+\left[\underline{C}_0^-\underline{l}\left( e^{\varphi_0}(\varphi_0)_\xi\int e^{-\varphi_0}d\xi+1\right)+B_0^-e^{\varphi_0}(\varphi_0)_\xi\right],\\
R^+=\underline{C}_0^+,\qquad R^-=\underline{C}_0^-,\qquad\phi=\varphi_0(\xi)+i\theta^+\theta^-(\varphi_0(\xi))_\xi,\qquad f=\psi(\xi),
\ea\label{sol62}
\ee
where $\varphi_0$ and $\psi$ are bosonic functions of the symmetry variable $\xi=x_--x_+$, while $\underline{C}_0^\pm,\underline{C}$ and $\underline{l}$ are arbitrary fermionic constants and $B_0^\pm$ are bosonic constants satisfying the algebraic constraint
\be
\underline{C}_0^+B_0^-+\underline{C}_0^-B_0^+=0.
\ee
For the solution (\ref{sol62}), the tangent vectors are linearly dependent, so the immersion defines curves instead of surfaces.

\paragraph{}\textbf{2.} For the subalgebra $\mathfrak{g}_{41}=\lbrace C_0+\epsilon P_+,\epsilon=\pm1\rbrace$, the orbit of the corresponding group of the SUSY GC equations (\ref{GC}) can be parametrized as follows
\be
\hspace{-2.5cm}\ba{l}
H=e^{\epsilon x_+}[h_0(x_-)+\theta^+\theta^-h_1(x_-)],\\
Q^\pm=e^{\epsilon x_+}[q^\pm_0(x_-)+\theta^+\theta^-q^\pm_1(x_-)],\\
R^\pm=r_0^\pm(x_-)+\theta^+\theta^-r_1^\pm(x_-),\\
\phi=\varphi_0(x_-)+\theta^+\theta^-\varphi_1(x_-),\quad f=e^{-2\epsilon x_+}\psi(x_-),
\ea
\ee
where the bosonic symmetry variable is $x_-$ and the fermionic symmetry variables are $\theta^+$ and $\theta^-$. An invariant solution of the SUSY GC equations (\ref{GC}) takes the form
\be\hspace{-2.5cm}
\ba{l}
H=2i\epsilon e^{\epsilon x_+-\varphi_0}\left[\underline{C}_0^+E_1-\frac{\epsilon}{E_0}\underline{C}_0^+(A_0E_1-A_1E_0)x_-+\theta^+\theta^-A_0\underline{C}_0^+\right],\\
Q^+=\underline{C}_0^+e^{\epsilon x_+}(E_0+\theta^+\theta^-E_1),\\
Q^-=\underline{C}_0^+e^{\epsilon x_+}(A_0+\theta^+\theta^-A_1),\\
R^+=\underline{C}_0^+,\quad R^-=\underline{C}_0^+,\quad\phi=\varphi_0(x_-)+\theta^+\theta^-\varphi_1(x_-), \quad f=e^{-2\epsilon x_+}\psi(x_-),
\ea\label{sol63}
\ee
where the bosonic functions $\varphi_0$ and $\varphi_1$ satisfy the conditions
\be
E_0\underline{C}_0^+\varphi_0=\underline{C}-\epsilon A_0\underline{C}_0^+x_-,\qquad \underline{C}_0^+\varphi_1=\frac{\epsilon}{E_0^2}\underline{C}_0^+(A_0E_1-A_1E_0)x_-,
\ee
respectively. Here $\psi$ is an arbitrary bosonic function of $x_-$ and $A_0,A_1,E_0$ and $E_1$ are arbitrary bosonic constants, while $\underline{C}_0^+$ and $\underline{C}$ are arbitrary fermionic constants.

The first and second fundamental forms of the surface $S$ (\ref{sol63}) are given by
\be
\hspace{-1.5cm}\ba{l}
I=e^{\varphi_0-2\epsilon x_+}\psi d_+d_-(1+\theta^+\theta^-\varphi_1),\\
I\hspace{-0.1cm}I=\underline{C}_0^+e^{-\epsilon x_+}\psi\left[d_+^2(E_0+\theta^+\theta^-E_1)+d_-^2(A_0+\theta^+\theta^-A_1)\right.\\
\hspace{1cm}+\left.2i\epsilon d_+d_-\left(E_1-\frac{\epsilon}{E_0}(A_0E_1-A_1E_0)x_-\right)\right.\\
\hspace{1cm}\left.+2i\epsilon d_+d_-\left(\theta^+\theta^-\left[A_0\underline{C}_0^++E_1\varphi_1-\frac{\epsilon}{E_0}(A_0E_1-A_1E_0)x_-\varphi_1\right]\right)\right].
\ea
\ee
The Gaussian curvature (\ref{Gaussian}) takes the form
\be
\mathcal{K}=0.
\ee

\paragraph{}\textbf{3.} For the subalgebra $\mathfrak{g}_{35}=\lbrace K_1+aK_0+bC_0,a\neq0,b\neq0\rbrace$, we obtain the following parametrization of the orbit of the corresponding group of the SUSY GC equations (\ref{GC})
\begin{equation*}\hspace{-2.5cm}\ba{ll}
H=(x_+)^{(a-b)/2}[h_0(x_-)+\eta\theta^-h_1(x_-)], & R^+=(x_+)^{-1/2}[r_0^+(x_-)+\eta\theta^-r_1^+(x_-)], \\
Q^+=(x_+)^{-(a+b+2)/2}[q_0^+(x_-)+\eta\theta^-q_1^+(x_-)], & R^-=r^-_0(x_-)+\eta\theta^-r_1^-(x_-),\\
Q^-=(x_+)^{-(a+b)/2}[q^-_0(x_-)+\eta\theta^-q^-_1(x_-)], & \phi=\varphi_0(x_-)+\eta\theta^-\varphi_1(x_-)-\frac{2a+1}{2}\ln x_+,\\
f=(x_+)^b\psi(x_-), & 
\ea\end{equation*}
where the bosonic symmetry variable is $x_-$ and the fermionic symmetry variables are $\eta=(x_+)^{-1/2}\theta^+$ and $\theta^-$. A corresponding invariant solution of the SUSY GC equations (\ref{GC}) has the form
\be
\hspace{-2.5cm}\ba{l}
H=(x_+)^{(a-b)/2}e^{A_0(a-b)x_-/2E_0}\left[\underline{C}+i(x_+)^{-1/2}\theta^+\theta^-(a-b+1)A_0\underline{C}_0^+e^{A_0x_-/2E_0}\right],\\
Q^+=\underline{C}_0^+(x_+)^{-(a+b+2)/2}\left[E_0+(x_+)^{-1/2}\theta^+\theta^-E_1\right],\quad R^+=\underline{C}_0^+(x_+)^{-1/2}, \\
Q^-=A_0\underline{C}_0^+(x_+)^{-(a+b)/2}\left[1+(x_+)^{-1/2}\theta^+\theta^-\frac{E_1}{E_0}\right],\qquad f=(x_+)^b\psi(x_-),\\
R^-=\underline{C}_0^+,\quad\phi=\frac{A_0}{2E_0}(b-a-1)x_-+(x_+)^{-1/2}\theta^+\theta^-\varphi_1(x_-)-\frac{2a+1}{2}\ln x_+,
\ea\label{sol64}
\ee
where the bosonic function $\varphi_1$ of $x_-$ satisfies the constraint
\be
\underline{C}_0^+\varphi_1=\frac{E_1}{E_0}\underline{C}_0^++i\frac{(a-b)}{4E_0}\underline{C}e^{-A_0x_-/2E_0},
\ee
and where $\psi$ is an arbitrary bosonic function $x_-$. Here $\underline{C}_0^+$ and $\underline{C}$ are arbitrary fermionic constants while $E_0,E_1$ and $A_0$ are arbitrary bosonic constants.

The first and second fundamental forms of the surface $S$ (\ref{sol64}) are given by
\begin{equation*}
\hspace{-2.5cm}\ba{l}
I=(x_+)^{(2b-2a-1)/2}\exp\left(\frac{A_0}{2E_0}(b-a-1)x_-\right)\psi d_+d_-(1+(x_+)^{-1/2}\theta^+\theta^-\varphi_1),\\
I\hspace{-0.1cm}I=(x_+)^{(b-a)/2}\psi\left[\underline{C}_0^+(x_+)^{-1}d_+^2(E_0+(x_+)^{-1/2}\theta^+\theta^- E_1)\right.\\
+A_0\underline{C}_0^+d_-^2(1+(x_+)^{-1/2}\theta^+\theta^-E_1/E_0)\\
+\left.(x_+)^{-1/2}e^{A_0x_-/2E_0}d_+d_-\left(\underline{C}+(x_+)^{-1/2}\theta^+\theta^-\left[i\underline{C}\varphi_1+(a-b+1)A_0\underline{C}_0^+e^{A_0x_-/2E_0}\right]\right)\right].
\ea
\end{equation*}
The Gaussian curvature (\ref{Gaussian}) takes the form
\be
\mathcal{K}=0.
\ee
Note that in our SUSY adaption of the classical geometric interpretation of surfaces in $\mathbb{R}^3$, the surfaces obtained in the three examples are composed of planar points or parabolic points.

\section{Conclusions}
In this paper, we have formulated a fermionic SUSY extension of the GW and GC equation ((\ref{GW}) and (\ref{GC}) respectively) for conformally parametrized surfaces immersed in a Grassmann superspace $\mathbb{R}^{(1,1\vert2)}$. It is interesting and significant to note that the obtained SUSY GW equations (\ref{GW}) and GC equations (\ref{GC}) for a SUSY version of a moving frame resemble the form of the classical equations (\ref{claGW}) and (\ref{claGC}), respectively. The SUSY extension of the GW equations (\ref{GW}) is obtained in the Bianchi form. The zero curvature condition (\ref{ZCC}) for the fermionic SUSY GC equations (\ref{GC0}) with the bosonic constant $f$ differs from its classical counterpart (\ref{claZCC}) in that it involves an anticommutator instead of a commutator. In addition, the signs of the last two terms change. The form of the zero-curvature condition (\ref{ZCC}) for the fermionic SUSY GC equations (\ref{GC0}) differs from that for the previously established bosonic SUSY extension of the GC equations, as well as from the bosonic ZCC for the SUSY sine-Gordon equation (\ref{sG}). It should be noted that equation (\ref{ZCC}) does not involve the matrix $E$ given by (\ref{E}). The SUSY sine-Gordon equation (\ref{sG}) does not determine a conformal parametrization of a surface. Hence, the surfaces associated with the SUSY sine-Gordon equation (\ref{sG}) and the link between the SUSY sine-Gordon model and the SUSY GW and GC equations require a separate investigation similar to the one performed in this paper.

The symmetries found for the obtained SUSY GC equation (\ref{GC}) include four bosonic dilations $C_0,K_0,K_1$ and $K_2$, two bosonic translations $P_+$ and $P_-$, and one fermionic shift $W$ together with the supersymmetry operators $J_+$ and $J_-$. This is in contrast with the symmetry algebra of the classical case, which is infinite-dimensional, and whose largest finite-dimensional subalgebra (\ref{clagen}) contains three dilations $e_0,e_3$ and $e_4$, two translations $e_1$ and $e_2$, and two conformal transformations $e_5$ and $e_6$. Also, the classical Lie symmetry algebra $A$ contains a center $e_0$, while its SUSY counterpart does not. The classification list (by conjugacy classes) of the one-dimensional subalgebras of the Lie symmetry algebra $\mathfrak{g}$ for the fermionic SUSY extension of the GC equations (\ref{GC}) includes $199$ subalgebras, which is different from both the equivalent classification for the classical model (which has $16$ subalgebras) and that for the bosonic SUSY extension of this model (which has 99 subalgebras) \cite{BGH}.

The first and second fundamental forms for SUSY conformally parametrized surfaces were established for the fermionic SUSY extension of the GC equations (\ref{GC}). These fundamental forms (\ref{I}) and (\ref{II}) differ in their signs from the classical case (equations (2.7) and (2.8) in \cite{BGH}). Also, we have established an analogue of the Bonnet Theorem for the fermionic SUSY GC equations (\ref{GC}). Three examples of solutions of the fermionic SUSY GC equations (\ref{GC}) are presented. For all three case, we found that the Gaussian curvature $K$ vanishes. However, the mean curvature $H$ is not zero. The relation $K\leqslant H^2$, present in the classical case, loses its meaning in the fermionic SUSY case.

This research could be extended in several directions. It could be beneficial to compute an exhaustive list of all symmetries of the fermionic SUSY GC equations (\ref{GC}) and to compare them to the classical case. The computation of such a list would require the development of a computer algebra Lie symmetry application capable of handling both even and odd Grassmann variables. Another open problem to be considered is whether all integrable SUSY systems possess non-standard invariants. It would be worthwhile to verify whether the conjecture proposed in \cite{CGS},\cite{LST} extends to all integrable SUSY models. It could also be worth attempting to establish a SUSY analogue of Noether's Theorem in order to study the conservation laws of such SUSY models. Finally, it would be interesting to investigate how the integrable characteristics, such as Hamiltonian structure and conserved quantities manifest themselves in surfaces for the SUSY case. These subjects will be investigated in our future work.

\section*{Acknowledgements}
AMG's work was supported by a research grant from NSERC of Canada. SB acknowledges a doctoral fellowship provided by the FQRNT of the Gouvernement du Qu\'ebec. AJH wishes to acknowledge and thank the Mathematical Physics Laboratory of the Centre de Recherches Math\'ematiques for the opportunity to contribute to this research.

\section*{Appendix. Classification of the one-dimensionnal subalgebras of the Lie superalgebra (\ref{cla}).}
\begin{table}[h!]
\caption{Classification of the one-dimensional subalgebras of the symmetry superalgebra $\mathfrak{g}$ of the equations (\ref{GC}) into conjugacy classes.  Here $\epsilon=\pm1$, the parameters $a,b,c$ are non-zero bosonic constants, $\underline{\mu},\underline{\nu},\underline{\zeta}$ are non-zero fermionic constants.}
\begin{tabular}{|l|l||l|l|}
\hline No & Subalgebra & No & Subalgebra \\
\hline $\mathfrak{g}_1$ & $\lbrace K_1\rbrace$ & $\mathfrak{g}_{36}$ & $\lbrace K_1+aK_0+\underline{\zeta}W\rbrace$ \\
\hline$\mathfrak{g}_2$ & $\lbrace P_+\rbrace$ & $\mathfrak{g}_{37}$ & $\lbrace K_1+aC_0+\underline{\zeta}W\rbrace$ \\
\hline$\mathfrak{g}_3$ & $\lbrace \underline{\mu}J_+\rbrace$ & $\mathfrak{g}_{38}$ & $\lbrace K_1+aK_0+bC_0+\underline{\zeta}W\rbrace$ \\
\hline$\mathfrak{g}_4$ & $\lbrace P_++\underline{\mu}J_+\rbrace$ & $\mathfrak{g}_{39}$ & $\lbrace P_++\underline{\zeta}W\rbrace$ \\
\hline$\mathfrak{g}_5$ & $\lbrace K_2\rbrace$ & $\mathfrak{g}_{40}$ & $\lbrace K_0+\epsilon P_+\rbrace$ \\
\hline$\mathfrak{g}_6$ & $\lbrace P_-\rbrace$ & $\mathfrak{g}_{41}$ & $\lbrace C_0+\epsilon P_+\rbrace$ \\
\hline$\mathfrak{g}_7$ & $\lbrace \underline{\nu}J_-\rbrace$ & $\mathfrak{g}_{42}$ & $\lbrace K_0+aC_0+\epsilon P_+\rbrace$ \\
\hline$\mathfrak{g}_8$ & $\lbrace P_-+\underline{\nu}J_-\rbrace$ & $\mathfrak{g}_{43}$ & $\lbrace K_0+\epsilon P_++\underline{\zeta}W\rbrace$ \\
\hline$\mathfrak{g}_9$ & $\lbrace K_1+aK_2\rbrace$ & $\mathfrak{g}_{44}$ & $\lbrace C_0+\epsilon P_++\underline{\zeta}W\rbrace$ \\
\hline$\mathfrak{g}_{10}$ & $\lbrace K_1+\epsilon P_-\rbrace$ & $\mathfrak{g}_{45}$ & $\lbrace K_0+aC_0+\epsilon P_++\underline{\zeta}W\rbrace$ \\
\hline$\mathfrak{g}_{11}$ & $\lbrace K_1+\underline{\nu}J_-\rbrace$ & $\mathfrak{g}_{46}$ & $\lbrace \underline{\mu}J_++\underline{\zeta}W\rbrace$ \\
\hline$\mathfrak{g}_{12}$ & $\lbrace K_1+\epsilon P_-+\underline{\nu}J_-\rbrace$ & $\mathfrak{g}_{47}$ & $\lbrace K_0+\underline{\mu}J_+\rbrace$ \\
\hline$\mathfrak{g}_{13}$ & $\lbrace K_2+\epsilon P_+\rbrace$ & $\mathfrak{g}_{48}$ & $\lbrace C_0+\underline{\mu}J_+\rbrace$ \\
\hline$\mathfrak{g}_{14}$ & $\lbrace P_++\epsilon P_-\rbrace$ & $\mathfrak{g}_{49}$ & $\lbrace K_0+aC_0+\underline{\mu}J_+\rbrace$ \\
\hline$\mathfrak{g}_{15}$ & $\lbrace P_++\underline{\nu}J_-\rbrace$ & $\mathfrak{g}_{50}$ & $\lbrace K_0+\underline{\mu}J_++\underline{\zeta}W\rbrace$ \\
\hline$\mathfrak{g}_{16}$ & $\lbrace P_++\epsilon P_-+\underline{\nu}J_-\rbrace$ & $\mathfrak{g}_{51}$ & $\lbrace C_0+\underline{\mu}J_++\underline{\zeta}W\rbrace$ \\
\hline$\mathfrak{g}_{17}$ & $\lbrace K_2+\underline{\mu}J_+\rbrace$ & $\mathfrak{g}_{52}$ & $\lbrace K_0+aC_0+\underline{\mu}J_++\underline{\zeta}W\rbrace$ \\
\hline$\mathfrak{g}_{18}$ & $\lbrace P_-+\underline{\mu}J_+\rbrace$ & $\mathfrak{g}_{53}$ & $\lbrace P_++\underline{\mu}J_++\underline{\zeta}W\rbrace$ \\
\hline$\mathfrak{g}_{19}$ & $\lbrace \underline{\mu}J_++\underline{\nu}J_-\rbrace$ & $\mathfrak{g}_{54}$ & $\lbrace K_0+\epsilon P_++\underline{\mu}J_+\rbrace$ \\
\hline$\mathfrak{g}_{20}$ & $\lbrace P_-+\underline{\mu}J_++\underline{\nu}J_-\rbrace$ & $\mathfrak{g}_{55}$ & $\lbrace C_0+\epsilon P_++\underline{\mu}J_+\rbrace$ \\
\hline$\mathfrak{g}_{21}$ & $\lbrace K_2+\epsilon P_++\underline{\mu}J_+\rbrace$ & $\mathfrak{g}_{56}$ & $\lbrace K_0+aC_0+\epsilon P_++\underline{\mu}J_+\rbrace$ \\
\hline$\mathfrak{g}_{22}$ & $\lbrace P_++\epsilon P_-+\underline{\mu}J_+\rbrace$ & $\mathfrak{g}_{57}$ & $\lbrace K_0+\epsilon P_++\underline{\mu}J_++\underline{\zeta}W\rbrace$ \\
\hline$\mathfrak{g}_{23}$ & $\lbrace P_++\underline{\mu}J_++\underline{\nu}J_-\rbrace$ & $\mathfrak{g}_{58}$ & $\lbrace C_0+\epsilon P_++\underline{\mu}J_++\underline{\zeta}W\rbrace$ \\
\hline$\mathfrak{g}_{24}$ & $\lbrace P_++\epsilon P_-+\underline{\mu}J_++\underline{\nu}J_-\rbrace$ & $\mathfrak{g}_{59}$ & $\lbrace K_0+aC_0+\epsilon P_++\underline{\mu}J_++\underline{\zeta}W\rbrace$ \\
\hline$\mathfrak{g}_{25}$ & $\lbrace \underline{\zeta}W\rbrace$ & $\mathfrak{g}_{60}$ & $\lbrace K_2+\underline{\zeta}W\rbrace$ \\
\hline$\mathfrak{g}_{26}$ & $\lbrace K_0\rbrace$ & $\mathfrak{g}_{61}$ & $\lbrace K_2+aK_0\rbrace$ \\
\hline$\mathfrak{g}_{27}$ & $\lbrace C_0\rbrace$ & $\mathfrak{g}_{62}$ & $\lbrace K_2+aC_0\rbrace$ \\
\hline$\mathfrak{g}_{28}$ & $\lbrace K_0+aC_0\rbrace$ & $\mathfrak{g}_{63}$ & $\lbrace K_2+aK_0+bC_0\rbrace$ \\
\hline$\mathfrak{g}_{29}$ & $\lbrace K_0+\underline{\zeta}W\rbrace$ & $\mathfrak{g}_{64}$ & $\lbrace K_2+aK_0+\underline{\zeta}W\rbrace$ \\
\hline$\mathfrak{g}_{30}$ & $\lbrace C_0+\underline{\zeta}W\rbrace$ & $\mathfrak{g}_{65}$ & $\lbrace K_2+aC_0+\underline{\zeta}W\rbrace$ \\
\hline$\mathfrak{g}_{31}$ & $\lbrace K_0+aC_0+\underline{\zeta}W\rbrace$ & $\mathfrak{g}_{66}$ & $\lbrace K_2+aK_0+bC_0+\underline{\zeta}W\rbrace$ \\
\hline$\mathfrak{g}_{32}$ & $\lbrace K_1+\underline{\zeta}W\rbrace$ & $\mathfrak{g}_{67}$ & $\lbrace P_-+\underline{\zeta}W\rbrace$ \\
\hline$\mathfrak{g}_{33}$ & $\lbrace K_1+aK_0\rbrace$ & $\mathfrak{g}_{68}$ & $\lbrace K_0+\epsilon P_-\rbrace$ \\
\hline$\mathfrak{g}_{34}$ & $\lbrace K_1+aC_0\rbrace$ & $\mathfrak{g}_{69}$ & $\lbrace C_0+\epsilon P_-\rbrace$ \\
\hline$\mathfrak{g}_{35}$ & $\lbrace K_1+aK_0+bC_0\rbrace$ & $\mathfrak{g}_{70}$ & $\lbrace K_0+aC_0+\epsilon P_-\rbrace$ \\
\hline
\end{tabular}
\centering
\end{table}
\setcounter{table}{2}
\begin{table}
\caption{(Continued)}
\begin{tabular}{|l|l||l|l|}
\hline No & Subalgebra & No & Subalgebra \\
\hline $\mathfrak{g}_{71}$ & $\lbrace K_0+\epsilon P_-+\underline{\zeta}W\rbrace$ & $\mathfrak{g}_{108}$ & $\lbrace K_0+aC_0+bK_1+\underline{\nu}J_-+\underline{\zeta}W\rbrace$ \\
\hline$\mathfrak{g}_{72}$ & $\lbrace C_0+\epsilon P_-+\underline{\zeta}W\rbrace$ & $\mathfrak{g}_{109}$ & $\lbrace K_1+\epsilon P_-+\underline{\nu}J_-+\underline{\zeta}W\rbrace$ \\
\hline$\mathfrak{g}_{73}$ & $\lbrace K_0+aC_0+\epsilon P_-+\underline{\zeta}W\rbrace$ & $\mathfrak{g}_{110}$ & $\lbrace K_0+aK_1+\epsilon P_-+\underline{\nu}J_-\rbrace$ \\
\hline$\mathfrak{g}_{74}$ & $\lbrace \underline{\nu}J_-+\underline{\zeta}W\rbrace$ & $\mathfrak{g}_{111}$ & $\lbrace C_0+aK_1+\epsilon P_-+\underline{\nu}J_-\rbrace$ \\
\hline$\mathfrak{g}_{75}$ & $\lbrace K_0+\underline{\nu}J_-\rbrace$ & $\mathfrak{g}_{112}$ & $\lbrace K_0+aC_0+bK_1+\epsilon P_-+\underline{\nu}J_-\rbrace$ \\
\hline$\mathfrak{g}_{76}$ & $\lbrace C_0+\underline{\nu}J_-\rbrace$ & $\mathfrak{g}_{113}$ & $\lbrace K_0+aK_1+\epsilon P_-+\underline{\nu}J_-+\underline{\zeta}W\rbrace$ \\
\hline$\mathfrak{g}_{77}$ & $\lbrace K_0+aC_0+\underline{\nu}J_-\rbrace$ & $\mathfrak{g}_{114}$ & $\lbrace C_0+aK_1+\epsilon P_-+\underline{\nu}J_-+\underline{\zeta}W\rbrace$ \\
\hline$\mathfrak{g}_{78}$ & $\lbrace K_0+\underline{\nu}J_-+\underline{\zeta}W\rbrace$ & $\mathfrak{g}_{115}$ & $\lbrace K_0+aC_0+bK_1+\epsilon P_-+\underline{\nu}J_-+\underline{\zeta}W\rbrace$ \\
\hline$\mathfrak{g}_{79}$ & $\lbrace C_0+\underline{\nu}J_-+\underline{\zeta}W\rbrace$ & $\mathfrak{g}_{116}$ & $\lbrace K_2+\epsilon P_++\underline{\zeta}W\rbrace$ \\
\hline$\mathfrak{g}_{80}$ & $\lbrace K_0+aC_0+\underline{\nu}J_-+\underline{\zeta}W\rbrace$ & $\mathfrak{g}_{117}$ & $\lbrace K_0+aK_2+\epsilon P_+\rbrace$ \\
\hline$\mathfrak{g}_{81}$ & $\lbrace P_-+\underline{\nu}J_-+\underline{\zeta}W\rbrace$ & $\mathfrak{g}_{118}$ & $\lbrace C_0+aK_2+\epsilon P_+\rbrace$ \\
\hline$\mathfrak{g}_{82}$ & $\lbrace K_0+\epsilon P_-+\underline{\nu}J_-\rbrace$ & $\mathfrak{g}_{119}$ & $\lbrace K_0+aC_0+bK_2+\epsilon P_+\rbrace$ \\
\hline$\mathfrak{g}_{83}$ & $\lbrace C_0+\epsilon P_-+\underline{\nu}J_-\rbrace$ & $\mathfrak{g}_{120}$ & $\lbrace K_0+aK_2+\epsilon P_++\underline{\zeta}W\rbrace$ \\
\hline$\mathfrak{g}_{84}$ & $\lbrace K_0+aC_0++\epsilon P_-\underline{\nu}J_-\rbrace$ & $\mathfrak{g}_{121}$ & $\lbrace C_0+aK_2+\epsilon P_++\underline{\zeta}W\rbrace$ \\
\hline$\mathfrak{g}_{85}$ & $\lbrace K_0+\epsilon P_-+\underline{\nu}J_-+\underline{\zeta}W\rbrace$ & $\mathfrak{g}_{122}$ & $\lbrace K_0+aC_0+bK_2+\epsilon P_++\underline{\zeta}W\rbrace$ \\
\hline$\mathfrak{g}_{86}$ & $\lbrace C_0+\epsilon P_-+\underline{\nu}J_-+\underline{\zeta}W\rbrace$ & $\mathfrak{g}_{123}$ & $\lbrace P_++\epsilon P_-+\underline{\zeta}W\rbrace$ \\
\hline$\mathfrak{g}_{87}$ & $\lbrace K_0+aC_0+\epsilon P_-+\underline{\nu}J_-+\underline{\zeta}W\rbrace$ & $\mathfrak{g}_{124}$ & $\lbrace P_++\epsilon P_-+aK_0\rbrace$ \\
\hline$\mathfrak{g}_{88}$ & $\lbrace K_1+aK_2+\underline{\zeta}W\rbrace$ & $\mathfrak{g}_{125}$ & $\lbrace P_++\epsilon P_-+aC_0\rbrace$ \\
\hline$\mathfrak{g}_{89}$ & $\lbrace K_0+aK_1+bK_2\rbrace$ & $\mathfrak{g}_{126}$ & $\lbrace P_++\epsilon P_-+aK_0+bC_0\rbrace$ \\
\hline$\mathfrak{g}_{90}$ & $\lbrace C_0+aK_1+bK_2\rbrace$ & $\mathfrak{g}_{127}$ & $\lbrace P_++\epsilon P_-+aK_0+\underline{\zeta}W\rbrace$ \\
\hline$\mathfrak{g}_{91}$ & $\lbrace K_0+aC_0+bK_1+cK_2\rbrace$ & $\mathfrak{g}_{128}$ & $\lbrace P_++\epsilon P_-+aC_0+\underline{\zeta}W\rbrace$ \\
\hline$\mathfrak{g}_{92}$ & $\lbrace K_0+aK_1+bK_2+\underline{\zeta}W\rbrace$ & $\mathfrak{g}_{129}$ & $\lbrace P_++\epsilon P_-+aK_0+bC_0\underline{\zeta}W\rbrace$ \\
\hline$\mathfrak{g}_{93}$ & $\lbrace C_0+aK_1+bK_2+\underline{\zeta}W\rbrace$ & $\mathfrak{g}_{130}$ & $\lbrace  P_++\underline{\nu}J_-+\underline{\zeta}W\rbrace$ \\
\hline$\mathfrak{g}_{94}$ & $\lbrace K_0+aC_0+bK_1+cK_2+\underline{\zeta}W\rbrace$ & $\mathfrak{g}_{131}$ & $\lbrace K_0+\epsilon P_++\underline{\nu}J_-\rbrace$ \\
\hline$\mathfrak{g}_{95}$ & $\lbrace K_1+\epsilon P_-+\underline{\zeta}W\rbrace$ & $\mathfrak{g}_{132}$ & $\lbrace C_0+\epsilon P_++\underline{\nu}J_-\rbrace$ \\
\hline$\mathfrak{g}_{96}$ & $\lbrace K_0+aK_1+\epsilon P_-\rbrace$ & $\mathfrak{g}_{133}$ & $\lbrace K_0+aC_0+\epsilon P_++\underline{\nu}J_-\rbrace$ \\
\hline$\mathfrak{g}_{97}$ & $\lbrace C_0+aK_1+\epsilon P_-\rbrace$ & $\mathfrak{g}_{134}$ & $\lbrace K_0+\epsilon P_++\underline{\nu}J_-+\underline{\zeta}W\rbrace$ \\
\hline$\mathfrak{g}_{98}$ & $\lbrace K_0+aC_0+bK_1+\epsilon P_-\rbrace$ & $\mathfrak{g}_{135}$ & $\lbrace C_0+\epsilon P_++\underline{\nu}J_-+\underline{\zeta}W\rbrace$ \\
\hline$\mathfrak{g}_{99}$ & $\lbrace K_0+aK_1+\epsilon P_-+\underline{\zeta}W\rbrace$ & $\mathfrak{g}_{136}$ & $\lbrace K_0+aC_0+\epsilon P_++\underline{\nu}J_-+\underline{\zeta}W\rbrace$ \\
\hline$\mathfrak{g}_{100}$ & $\lbrace C_0+aK_1+\epsilon P_-+\underline{\zeta}W\rbrace$ & $\mathfrak{g}_{137}$ & $\lbrace P_++\epsilon P_-+\underline{\nu}J_-+\underline{\zeta}W\rbrace$ \\
\hline$\mathfrak{g}_{101}$ & $\lbrace K_0+aC_0+bK_1+\epsilon P_-+\underline{\zeta}W\rbrace$ & $\mathfrak{g}_{138}$ & $\lbrace P_++\epsilon P_-+aK_0+\underline{\nu}J_-\rbrace$ \\
\hline$\mathfrak{g}_{102}$ & $\lbrace K_1+\underline{\nu}J_-+\underline{\zeta}W\rbrace$ & $\mathfrak{g}_{139}$ & $\lbrace P_++\epsilon P_-+aC_0+\underline{\nu}J_-\rbrace$ \\
\hline$\mathfrak{g}_{103}$ & $\lbrace K_0+aK_1+\underline{\nu}J_-\rbrace$ & $\mathfrak{g}_{140}$ & $\lbrace P_++\epsilon P_-+aK_0+bC_0+\underline{\nu}J_-\rbrace$ \\
\hline$\mathfrak{g}_{104}$ & $\lbrace C_0+aK_1+\underline{\nu}J_-\rbrace$ & $\mathfrak{g}_{141}$ & $\lbrace P_++\epsilon P_-+aK_0+\underline{\nu}J_-+\underline{\zeta}W\rbrace$ \\
\hline$\mathfrak{g}_{105}$ & $\lbrace K_0+aC_0+bK_1+\underline{\nu}J_-\rbrace$ & $\mathfrak{g}_{142}$ & $\lbrace P_++\epsilon P_-+aC_0+\underline{\nu}J_-+\underline{\zeta}W\rbrace$ \\
\hline$\mathfrak{g}_{106}$ & $\lbrace K_0+aK_1+\underline{\nu}J_-+\underline{\zeta}W\rbrace$ & $\mathfrak{g}_{143}$ & $\lbrace P_++\epsilon P_-+aK_0+bC_0+\underline{\nu}J_-+\underline{\zeta}W\rbrace$ \\
\hline$\mathfrak{g}_{107}$ & $\lbrace C_0+aK_1+\underline{\nu}J_-+\underline{\zeta}W\rbrace$ & $\mathfrak{g}_{144}$ & $\lbrace K_2+\underline{\mu}J_++\underline{\zeta}W\rbrace$ \\
\hline
\end{tabular}
\centering
\end{table}
\setcounter{table}{2}
\begin{table}
\caption{(Continued)}
\begin{tabular}{|l|l||l|l|}
\hline No & Subalgebra & No & Subalgebra\\
\hline$\mathfrak{g}_{145}$ & $\lbrace K_0+aK_2+\underline{\mu}J_+\rbrace$ & $\mathfrak{g}_{173}$ & $\lbrace K_0+aK_2+\epsilon P_++\underline{\mu}J_+\rbrace$ \\
\hline$\mathfrak{g}_{146}$ & $\lbrace C_0+aK_2+\underline{\mu}J_+\rbrace$ & $\mathfrak{g}_{174}$ & $\lbrace C_0+aK_2+\epsilon P_++\underline{\mu}J_+\rbrace$ \\
\hline$\mathfrak{g}_{147}$ & $\lbrace K_0+aC_0+bK_2+\underline{\mu}J_+\rbrace$ & $\mathfrak{g}_{175}$ & $\lbrace K_0+aC_0+bK_2+\epsilon P_++\underline{\mu}J_+\rbrace$ \\
\hline$\mathfrak{g}_{148}$ & $\lbrace K_0+aK_2+\underline{\mu}J_++\underline{\zeta}W\rbrace$ & $\mathfrak{g}_{176}$ & $\lbrace K_0+aK_2+\epsilon P_++\underline{\mu}J_++\underline{\zeta}W\rbrace$ \\
\hline$\mathfrak{g}_{149}$ & $\lbrace C_0+aK_2+\underline{\mu}J_++\underline{\zeta}W\rbrace$ & $\mathfrak{g}_{177}$ & $\lbrace C_0+aK_2+\epsilon P_++\underline{\mu}J_++\underline{\zeta}W\rbrace$ \\
\hline$\mathfrak{g}_{150}$ & $\lbrace K_0+aC_0+bK_2+\underline{\mu}J_++\underline{\zeta}W\rbrace$ & $\mathfrak{g}_{178}$ & $\lbrace K_0+aC_0+bK_2+\epsilon P_++\underline{\mu}J_++\underline{\zeta}W\rbrace$ \\
\hline$\mathfrak{g}_{151}$ & $\lbrace P_-+\underline{\mu}J_++\underline{\zeta}W\rbrace$ & $\mathfrak{g}_{179}$ & $\lbrace P_++\epsilon P_-+\underline{\mu}J_++\underline{\zeta}W\rbrace$ \\
\hline$\mathfrak{g}_{152}$ & $\lbrace K_0+\epsilon P_-+\underline{\mu}J_+\rbrace$ & $\mathfrak{g}_{180}$ & $\lbrace P_++\epsilon P_-+aK_0+\underline{\mu}J_+\rbrace$ \\
\hline$\mathfrak{g}_{153}$ & $\lbrace C_0+\epsilon P_-+\underline{\mu}J_+\rbrace$ & $\mathfrak{g}_{181}$ & $\lbrace P_++\epsilon P_-+aC_0+\underline{\mu}J_+\rbrace$ \\
\hline$\mathfrak{g}_{154}$ & $\lbrace K_0+aC_0+\epsilon P_-+\underline{\mu}J_+\rbrace$ & $\mathfrak{g}_{182}$ & $\lbrace P_++\epsilon P_-+aK_0+bC_0+\underline{\mu}J_+\rbrace$ \\
\hline$\mathfrak{g}_{155}$ & $\lbrace K_0+\epsilon P_-+\underline{\mu}J_++\underline{\zeta}W\rbrace$ & $\mathfrak{g}_{183}$ & $\lbrace P_++\epsilon P_-+aK_0+\underline{\mu}J_++\underline{\zeta}W\rbrace$ \\
\hline$\mathfrak{g}_{156}$ & $\lbrace C_0+\epsilon P_-+\underline{\mu}J_++\underline{\zeta}W\rbrace$ & $\mathfrak{g}_{184}$ & $\lbrace P_++\epsilon P_-+aC_0+\underline{\mu}J_++\underline{\zeta}W\rbrace$ \\
\hline$\mathfrak{g}_{157}$ & $\lbrace K_0+aC_0+\epsilon P_-+\underline{\mu}J_++\underline{\zeta}W\rbrace$ & $\mathfrak{g}_{185}$ & $\lbrace P_++\epsilon P_-+aK_0+bC_0+\underline{\mu}J_++\underline{\zeta}W\rbrace$ \\
\hline$\mathfrak{g}_{158}$ & $\lbrace \underline{\mu}J_++\underline{\nu}J_-+\underline{\zeta}W\rbrace$ & $\mathfrak{g}_{186}$ & $\lbrace P_++\underline{\mu}J_++\underline{\nu}J_-+\underline{\zeta}W\rbrace$ \\
\hline$\mathfrak{g}_{159}$ & $\lbrace K_0+\underline{\mu}J_++\underline{\nu}J_-\rbrace$ & $\mathfrak{g}_{187}$ & $\lbrace K_0+\epsilon P_++\underline{\mu}J_++\underline{\nu}J_-\rbrace$ \\
\hline$\mathfrak{g}_{160}$ & $\lbrace C_0+\underline{\mu}J_++\underline{\nu}J_-\rbrace$ & $\mathfrak{g}_{188}$ & $\lbrace C_0+\epsilon P_++\underline{\mu}J_++\underline{\nu}J_-\rbrace$ \\
\hline$\mathfrak{g}_{161}$ & $\lbrace K_0+aC_0+\underline{\mu}J_++\underline{\nu}J_-\rbrace$ & $\mathfrak{g}_{189}$ & $\lbrace K_0+aC_0+\epsilon P_++\underline{\mu}J_++\underline{\nu}J_-\rbrace$ \\
\hline$\mathfrak{g}_{162}$ & $\lbrace K_0+\underline{\mu}J_++\underline{\nu}J_-+\underline{\zeta}W\rbrace$ & $\mathfrak{g}_{190}$ & $\lbrace K_0+\epsilon P_++\underline{\mu}J_++\underline{\nu}J_-+\underline{\zeta}W\rbrace$ \\
\hline$\mathfrak{g}_{163}$ & $\lbrace C_0+\underline{\mu}J_++\underline{\nu}J_-+\underline{\zeta}W\rbrace$ & $\mathfrak{g}_{191}$ & $\lbrace C_0+\epsilon P_++\underline{\mu}J_++\underline{\nu}J_-+\underline{\zeta}W\rbrace$ \\
\hline$\mathfrak{g}_{164}$ & $\lbrace K_0+aC_0+\underline{\mu}J_++\underline{\nu}J_-+\underline{\zeta}W\rbrace$ & $\mathfrak{g}_{192}$ & $\lbrace K_0+aC_0+\epsilon P_++\underline{\mu}J_++\underline{\nu}J_-+\underline{\zeta}W\rbrace$ \\
\hline$\mathfrak{g}_{165}$ & $\lbrace P_-+\underline{\mu}J_++\underline{\nu}J_-+\underline{\zeta}W\rbrace$ & $\mathfrak{g}_{193}$ & $\lbrace P_++\epsilon P_-+\underline{\mu}J_++\underline{\nu}J_-+\underline{\zeta}W\rbrace$ \\
\hline$\mathfrak{g}_{166}$ & $\lbrace K_0+\epsilon P_-+\underline{\mu}J_++\underline{\nu}J_-\rbrace$ & $\mathfrak{g}_{194}$ & $\lbrace P_++\epsilon P_-+aK_0+\underline{\mu}J_++\underline{\nu}J_-\rbrace$ \\
\hline$\mathfrak{g}_{167}$ & $\lbrace C_0+\epsilon P_-+\underline{\mu}J_++\underline{\nu}J_-\rbrace$ & $\mathfrak{g}_{195}$ & $\lbrace P_++\epsilon P_-+aC_0+\underline{\mu}J_++\underline{\nu}J_-\rbrace$ \\
\hline$\mathfrak{g}_{168}$ & $\lbrace K_0+aC_0+\epsilon P_-+\underline{\mu}J_++\underline{\nu}J_-\rbrace$ & $\mathfrak{g}_{196}$ & $\lbrace P_++\epsilon P_-+aK_0+bC_0+\underline{\mu}J_++\underline{\nu}J_-\rbrace$ \\
\hline$\mathfrak{g}_{169}$ & $\lbrace K_0+\epsilon P_-+\underline{\mu}J_++\underline{\nu}J_-+\underline{\zeta}W\rbrace$ & $\mathfrak{g}_{197}$ & $\lbrace P_++\epsilon P_-+aK_0+\underline{\mu}J_++\underline{\nu}J_-+\underline{\zeta}W\rbrace$ \\
\hline$\mathfrak{g}_{170}$ & $\lbrace C_0+\epsilon P_-+\underline{\mu}J_++\underline{\nu}J_-+\underline{\zeta}W\rbrace$ & $\mathfrak{g}_{198}$ & $\lbrace P_++\epsilon P_-+aC_0+\underline{\mu}J_++\underline{\nu}J_-+\underline{\zeta}W\rbrace$ \\
\hline$\mathfrak{g}_{171}$ & $\lbrace K_0+aC_0+\epsilon P_-+\underline{\mu}J_++\underline{\nu}J_-+\underline{\zeta}W\rbrace$ & $\mathfrak{g}_{199}$ & $\lbrace P_++\epsilon P_-+aK_0+bC_0+\underline{\mu}J_++\underline{\nu}J_-+\underline{\zeta}W\rbrace$ \\
\hline$\mathfrak{g}_{172}$ & $\lbrace K_2+\epsilon P_++\underline{\mu}J_++\underline{\zeta}W\rbrace$ &  &  \\
\hline
\end{tabular}
\centering
\end{table}

\end{document}